\documentclass[aps,prl,nofootinbib,showpacs,twocolumn,groupedaddress,floatfix]{revtex4-1}

\pdfoutput=1
\usepackage{amsmath}
\usepackage{cancel}
\usepackage{float}
\usepackage{ifsym}
\usepackage{color}
\usepackage{latexsym}
\usepackage{latexsym}
\usepackage{dcolumn}
\usepackage{epsfig}
\usepackage{amsfonts}
\usepackage{amsmath}
\usepackage{bm}
\usepackage{graphicx}
\usepackage{hyperref} 

\newcommand{\bnull}{{\bm{0}}}
\newcommand{\bK}{{\bm{K}}}
\newcommand{\bQ}{{\bm{Q}}}

\newcommand{\bk}{{\bm{k}}}

\newcommand{\bp}{{\bm{p}}}
\newcommand{\bq}{{\bm{q}}}

\newcommand{\bx}{{\bm{x}}}

\newcommand{\al}{\alpha}
\newcommand{\ph}{\varphi}
\newcommand{\la}{\lambda}
\newcommand{\eps}{\epsilon}

\newcommand{\La}{\Lambda}
\newcommand{\om}{\omega}

\newcommand{\calD}{{\cal D}}

\newcommand{\calH}{{\cal H}}

\newcommand{\calK}{{\cal K}}

\newcommand{\calF}{{\cal F}}
\newcommand{\calM}{{\cal M}}

\newcommand{\calR}{{\cal R}}

\newcommand{\wh}{\widehat}

\newcommand{\entwederoder}[4]{\left\{\begin{array}{lll}#1,&&#2\\#3,&&#4\end{array}\right.}

\newcommand{\Keff}{{\calK_{\mathrm{eff}}}}                          
\newcommand{\Reff}{{\calR^{\mathrm{eff}}}}                          
\newcommand{\bKeff}{{\bK_{\mathrm{eff}}}}                          
\newcommand{\pieff}{{\pi_{\mathrm{eff}}}}                          
\newcommand{\deff}{{d_{\mathrm{eff}}}}                          

\begin{document}

\title{Fourier Monte Carlo Renormalization Group Approach to Crystalline Membranes}
\author{A.~Tr\"oster}
\email{andreas.troester@tuwien.ac.at}
\affiliation{Vienna University of Technology,
Institute of Material Chemistry, Getreidemarkt 9,
A-1060 Wien, Austria}
\date{\today}
 
\begin{abstract}
The computation of the critical exponent $\eta$  characterizing the universal elastic behavior of crystalline membranes in the flat phase 
continues to represent challenges to theorists as well as computer simulators that manifest themselves in a considerable spread of numerical 
results for $\eta$ published in the literature. We present new insight to this problem that results from combining Wilson's momentum shell renormalization group 
method with the power of modern computer simulations based on the Fourier Monte Carlo algorithm. After discussing the ideas and difficulties underlying 
this combined scheme, we present a calculation of the renormalization group flow of the 
effective 2d Young modulus for momentum shells of different thickness. Extrapolation to infinite shell thickness allows 
to produce results in reasonable agreement with those obtained by functional renormalization group or by 
Fourier Monte Carlo simulations in combination with finite size scaling.
Moreover, our new method allows for the first time to obtain a decent estimate for the value of the Wegner exponent 
$\om$ that determines the leading correction to scaling, which in turn allows to refine our numerical estimate for $\eta$  previously obtained
from precise finite size scaling data.
\end{abstract}

\pacs{64.60.De 05.10.Ln 46.70.Hg 05.70.Jk}

\maketitle

\section{Introduction}

The paradigm of the renormalization group (RG) is without doubt a cornerstone of modern theoretical physics with countless applications, and 
has been enormously influential in many areas of science beyond its origins rooted in high energy physics and statistical mechanics \cite{Wilson_RMP55_1983}.
Indeed, the abstract RG concept may be regarded as a cleverly organized successive divide-and-conquer strategy to deal with problems that
involve a large number of mutually coupled degrees of freedom. Yet, concrete applications of an RG scheme may superficially appear to look very different from one another.
In the present article we shall concentrate on Wilson's momentum shell RG (MSRG) approach to the field-theoretic formulation of critical phenomena at second order phase transitions
\cite{Zinn-Justin_QFTCP_2002}. 
The MSRG is certainly an invaluable conceptual tool both for abstract reasoning as well as in a first qualitative or even semi-quantitative analysis of a given problem.
In a nutshell, one writes the underlying Hamiltonian in terms of Fourier amplitudes $\tilde f(\bk)$ of the underlying fields. Imposing a wave vector cutoff $\La$,
one tries to identify an effective Hamiltonian as it would emerge after having integrated out all microscopic degrees of freedom that describe the physics of the system below scales
of size $1/\La$. In this effective Hamiltonian
only those couplings are kept that are regarded as important in the long wavelength limit, while the effect of all other couplings that are related to the eliminated short-ranged degrees 
of freedom is absorbed into an assumed renormalization of these surviving couplings.
The fact that the choice of the cutoff $\La$ is arbitrary suggest to iterate this prescription as follows. 
The effects of the ``fastest'' degrees of freedom which reside in a momentum shell $\La/b<|\bk|\le\La,\, b>1$ beneath the cutoff $\La$ are successively integrated out from
the partition function, which gives rise to a yet another set of modified coupling constants. On properly rescaling lengths and ``renormalizing'' the field, one derives a flow pattern in the 
space of coupling constants.
An analysis of the fixed points (FPs) of this flow then allows to explain the phenomenon of universality and to extract numerical values for the critical exponents.   
Unfortunately, concrete analytical implementations of this program usually rely on some type of perturbative approximation, and calculations frequently become intractable beyond one loop order.
Thus, for actual numerical calculations other approaches like the field-theoretic RG \cite{KleinertSchulteFrolinde_CP_2001} 
or the functional RG \cite{Kopietz2010} are preferred, or one resorts to
real space computer simulations in combination with finite size scaling (FSS) \cite{AmitMartinMayor_FRRGCP_2005,LandauBinder_MC_2009}. 

For the task of implementing Wilson's MSRG scheme in a simulation, real space MC approaches are obviously not very well suited.
On the other hand, our Fourier Monte Carlo algorithm (FMC) 
\cite{Troester_PRB76_2007,Troester_CSSCMP_2008,Troester_PRL100_2008,TroesterCPC179_2008,Troester_PRB_81_012406_2010} is tailor-made for this problem.
Recently we have demonstrated that it allows to follow the MSRG prescription step by step in simulation \cite{Troester_PRE_79_036707_2009,Troester_CPC182_2011}. 
This is quite appealing, as it eliminates the perturbative approximations 
and the underlying need for a ``small parameter'' from the concrete application of the MSRG, thus representing a truly nonperturbative implementation.
On the other hand, like in any humanly possible MSRG calculation, one is still forced to  project the calculated RG flow from its 
native infinite-dimensional coupling space to
a suitable low-dimensional subspace spanned by a finite number of effective coupling parameters.
Even though no perturbative approximation is involved, the presence of this inevitable projection, which amounts to ignoring the effects of 
the remaining directions in the space of coupling constants, may do substantial harm to the achieved numerical precision.    

Inspired by early analytic work by Bruce, Droz and Aharony \cite{BruceDrozAharony_JPCSSP7_1974}, 
subsequent work \cite{Troester_PRB81_125135_2010,Troester_CPC182_2011} indicates that
by optimizing the results with respect to the parameter $b$ that governs the thickness of the momentum shell (we will discuss below in more detail how this works)
MSRG can indeed be turned from a qualitatively to a quantitatively useful tool.
Up to date this has only been demonstrated for a particularly convenient model system, namely the long-ranged Ising model of
Fisher, Ma and Nickel \cite{FisherMaNickel_PRL_1972}.   
One purpose of the present paper is to test the ideas put forward in Refs.~\cite{Troester_PRB81_125135_2010,Troester_CPC182_2011} on a nontrivial real world problem, 
namely the elastic behavior of crystalline membranes
in the flat phase. It is well known \cite{NelsonPiranWeinberg_Membranes_1988} 
that the corresponding universal behavior of long wavelength fluctuations is governed by single exponent $\eta$.
However, as noted in Ref.~\cite{Troester_PRB_87_104112_2013},
a glance at the existing literature reveals a considerable spread of numerical results for $\eta$, obtained from a variety of analytical approaches like e.g.
self-consistent field theory \cite{LeDoussalRadzihovsky_PRL69_1992,Gazit_PRE80_041117_2009}, $\eps$-expansion \cite{AronovitzLubensky_PRL60_1988},
large d expansion \cite{DavidGuitter_EPL5_1988} and 
functional RG \cite{KownackiMouhanna_PRE79_040101_2009,Braghin_PRB82_035407_2010,Hasselmann_PRE83_031137_2011}, or from
simulation approaches derived in real space (see e.g.\ Refs.\ \cite{Bowick_JPF6_1321_1996,LosFasolino_PRB80_121405_2009}). 
In our own simulations
(Ref.~\cite{Troester_PRB_87_104112_2013}), which are based on our Fourier MC algorithm in combination with FSS,
we have noticed a rather strong influence of corrections to scaling, which indicates the importance of 
properly taking into account the role of RG-irrelevant couplings if one aims at high numerical precision.
Interestingly, up to date nobody seems to have succeeded in deriving a numerical estimate of the exponent $\om$ governing the
corrections to scaling (cf.~\cite{Wiese_PMAR_2000}) of a crystalline membrane in its flat phase. 
It is the second goal of the present paper to provide such a numerical estimate.

\section{A different view on standard MSRG}

As explained in the introduction, MSRG is a fairly standard method. Yet, for the convenience of the reader we will summarize 
the main steps, emphasizing those aspects that are particularly relevant to our present approach.
A MSRG transformation can be performed on an arbitrary Hamiltonian $\calH^{\La}_{\bK}[f]$ formulated in terms of the 
Fourier amplitudes $\tilde f(\bk)$ of a field $f(\bx)$ defined for wavevectors of moduli up to a chosen momentum space cutoff $\La$. 
The formal vector $\bK$ holds all ``coupling constants'' that are admissible for the symmetry constraints imposed on the
underlying system. Let $\calK$ denote the infinite-dimensional space of all such coupling vectors.   
One chooses a shell thickness parameter $b>1$ and splits the Fourier amplitudes $\tilde f(\bk)\equiv \tilde f_<(\bk)+\tilde f_>(\bk)$ 
into ``slow'' and ``fast'' contributions
\begin{eqnarray}
\tilde f_<(\bk)&=&\theta(\La/b-|\bk|)\tilde f(\bk)\,,\\
\tilde f_>(\bk)&=&\theta(|\bk|-\La/b)\tilde f(\bk)\,,
\end{eqnarray}
where $\theta$ denotes the Heaviside step function.
Functional integration over the fast modes 
\begin{eqnarray}
e^{-\tilde\calH^{\La/b}_{\tilde \bK}[f_<]}\equiv \int \calD f_>
e^{-\calH^\La_{\bK}[f_<+f_>]} 
\end{eqnarray}
then yields a new coarse-grained Hamiltonian for remaining slow modes
and induces a mapping $\bK\to\tilde \bK$.
This coarse-graining step is followed by a rescaling $\bk'=b\bk$ of ``momenta'' (i.e.~inverse length) scales
and restoration of the original cutoff $\La$.
Finally, one performs a renormalization 
\begin{eqnarray}
\tilde f_<(\bk'/b)=z(b,\bK)f'(\bk')  
\label{eqn:cdccdcccnnjn}
\end{eqnarray}
of field amplitudes with
\begin{eqnarray}
z(b,\bK)=b^{d-[f]-\frac{\eta[\bK]}{2}} \,.
\label{eqn:ccmkkmmkcncncd}
\end{eqnarray}
Here $d$ is the spatial dimension, $[f]$ is the canonical momentum dimension of $f$, and the so-called anomalous dimension $\eta[\bK]/2$ is a function which characterizes
the specific FP to be investigated (see below).
For the example of a coupling constant $C$ multiplying a monomial containing $n$ powers of the field $f$ and $p$ spatial derivatives in the effective Hamiltonian, it is 
straightforward to show that the coarse-grained coefficient $\tilde C$ undergoes a total rescaling 
\begin{eqnarray}
C'=b^{[C]-n\frac{\eta[\bk]}{2}}\tilde C  
\label{eqn:cnhcbjhcbhcbbqbqab}
\end{eqnarray}
where $p$ is implicitly accounted for in the canonical momentum dimension $[C]=d-n\cdot [f]-p$.
 
Consecutive application of these three steps induces a mapping $\tilde\bK\to\bK'$, which defines the RG
transformation $\calR_b:\calK\to\calK$.
The crux of the whole construction is the observation that as a result of the rescaling operation, the correlation lengths of systems at 
$\bK$ and $\bK'$ are related by $\xi[\bK']=\xi[\bK]/b$. At a FP $\bK^*=\calR_b(\bK^*)$ this leaves only the possibility of an infinite or zero correlation length.
Each such FP characterizes a different universality class of critical behavior, and  nontrivial behavior is, of course, found for infinite correlation length.    

In principle, the operation $\calR_b$ can be defined for any $b>1$, and satisfies the eponymous semi-group property 
\begin{eqnarray}
\calR_{b_1b_2}=\calR_{b_1}\circ\calR_{b_2},
\end{eqnarray}
which is paramount to the emergence of power laws that dominate the subsequent analysis as well as to the independence of the  
associated exponent values of the particular choice of the shell thickness parameter $b$. 
In the vicinity of $\bK^*$ where $\calR_b$ can be linearized,
most directions in the space $\calK$ turn out to be exponentially attractive (``irrelevant''), 
while typically only one or two are exponentially repulsive (``relevant''), and thus must be carefully tuned to ``reach'' the FP $\bK^*$
under successive iteration of $\calR_b$. Ultimately, this explains the observed universality of critical phenomena.
The RG flow resulting from the above scheme is defined in the infinite-dimensional coupling constant space $\calK$.
In practical calculations, one is nevertheless forced to limit ourselves  to working with \emph{effective} Hamiltonians, i.e.~Hamiltonians $\calH^{\La}_{\bKeff}[f]$ 
that are parametrized exclusively by coupling vectors $\bKeff\in\Keff$ taken from a low-dimensional linear subspace $\Keff\subset\calK$ of dimension, say, $\deff$,
spanned by the relevant and the least irrelevant directions w.r.t.~the FP $\bK^*$. 
In terms of suitably chosen coordinates in the space $\calK$, the projection $\pieff: \calK\to\Keff$ onto this finite-dimensional space assumes the form
\begin{eqnarray}
\pieff(K_1,K_2,\dots)=(K_1,K_2,\dots K_{\deff},0,0,\dots)\,.
\end{eqnarray} 
Except for trivial cases, $\Keff$ is not an invariant subspace under the action of $\calR_b$ i.e.~$\calR_b$ and $\pieff$ do not commute, because 
``new'' couplings are inevitably generated from a generic effective Hamiltonian under the coarse-graining operation, regardless of our ability
to perform the coarse graining operation exactly or by some approximate method. If the subspace $\Keff\subset\calK$ has been chosen properly,
the ``missing'' directions will only correspond to strongly irrelevant directions in coupling space, whose influence will be exponentially suppressed. 
In mathematical terms, any humanly possible MSRG calculation amounts to replacing the exact RG transformation $\calR$ by the effective transformation
\begin{eqnarray}
\Reff_b:=\pieff\circ\calR_b\circ\pieff  \,.
\end{eqnarray}
The crucial observation is, however, that, in contrast to $\calR_b$ the effective transformations $\Reff_b$ do \emph{not} strictly form a half-group, i.e.
\begin{eqnarray}
\Reff_{b_1b_2}\ne\Reff_{b_1}\circ\Reff_{b_2}\,,
\end{eqnarray}
since in the composite operation on the right side the additional irrelevant couplings generated by $\Reff_{b}$ will be ``lost'' in the subsequent application of $\pieff$.
Of course, nothing can prevent us from studying iterations of the map $\Reff_{b}$ in a manner similar to $\calR_{b}$.
The harm that the failure of $\Reff_{b}$ to close under composition causes to the subsequent analysis depends on the 
``production rate'' of coupling components generated during the coarse graining step that fall outside of $\Keff$, which in turn is controlled by the shell thickness parameter $b$.
In particular, the projection $\pieff(\bK^*)$ of the ``true'' infinite-dimensional FP $\bK^*$ of $\calR_{b}$ does generally \emph{not} produce a FP of $\Reff_{b}$.
Instead, the locations of FPs $\bKeff^*=\bKeff^*(b)\in\Keff$ of the transformations $\Reff_{b}$ will generally be \emph{$b$-dependent}. Furthermore, the same is true
for the numerical values of critical exponents calculated from a linearization of $\Reff_{b}$ around $\bKeff^*(b)$.

In summary, even though the results of the exact RG prescription in infinite-dimensional coupling space $\calK$ are guaranteed to be independent of the arbitrary parameter $b$,
the projection $\pieff$ to the low-dimensional space $\Keff$ introduces such a $b$-dependence that encodes the effects of the remaining irrelevant directions.
While this seems to look pathological at first sight, it actually allows to optimize the resulting calculation scheme by determining the value $b^*$ 
at which the drift of $\bKeff(b)$ becomes stationary. In this respect, our philosophy is similar to that of other approaches in which an arbitrary parameter is introduced
whose value would drop out of the results of exact theory but nevertheless may be used to optimize an approximated version. A nice example illustrating the power of such a
strategy is H.~Kleinert's ``variational perturbation theory'' \cite{Kleinert_PathIntegrals_2009}.  
However, it is very important to keep in mind that the present $b$-related ``pathologies'' are \emph{non-perturbative} in the sense that they do not originate from the use of any
perturbative approximation in evaluating the CG step, but purely arise from the necessity to limit ourselves to considering a finite number of couplings in a 
real world calculation. Amusingly, these effects are neither noticed in standard perturbative MSRG calculations, where 
it is extremely convenient to consider momentum shells that are infinitesimally thin, since in the limit 
$\Delta b:=b-1\to 0_+$ the appearing Feynman integrals are usually much easier to evaluate than for finite $\Delta b$,
nor in most popular real-space RG schemes where the value of $b$ is usually dictated by the decimation scheme chosen for the given lattice topology. 
In fact, it is difficult to find any papers that use momentum shells of finite thickness for anything beyond qualitative arguments.
One notable exception is the work of Bruce, Droz and Aharony \cite{BruceDrozAharony_JPCSSP7_1974}, who argued that the influence of irrelevant couplings in perturbative calculations of the
exponents of a standard short-ranged  Landau-Ginzburg (LG) model should be greatly diminished in the limit of large $b$.
And indeed, notice that $b^*\to\infty$ and $b^*\to1$ are the only values of $b^*$ that allow to reconcile the expected $b$-dependent features discussed above with the validity of the usual
semi-group property $\calR_{{b^*}^2}=\calR_{{b^*}}\circ\calR_{{b^*}}$.

Our recently developed FMC method is non-perturbative by definition and necessarily uses momentum shells of finite thickness, since 
our simulations are done for a finite lattice of linear size $L$ with lattice constant $a=1$, which implies a minimum spacing of
$\Delta k_i=2\pi/L$ between components of adjacent wavevectors. Thus, it is perfectly suited to study the $b$-dependence of $\Reff_{b}$ and
check the predictions of Bruce et al.~that had been derived with the use of the $\eps$-expansion.
Of course, due to the discrete nature of the Brillouin zones of our finite systems neither the limit $b\to1$ nor the limit  $b\to\infty$ 
are directly accessible, but we can monitor or even try to extrapolate the behavior of the corresponding observables towards these limits.

For the purpose of putting our ideas to the test, the short range LG model used in Ref.~\cite{BruceDrozAharony_JPCSSP7_1974} is not very suitable 
in view of the numerical smallness of its exponent $\eta=\eps^2/54+O(\eps^3)$. Instead, in Ref.~\cite{Troester_PRB81_125135_2010} we considered 
the long-range generalization of the LG model introduced by Fisher, Ma and Nickel in Ref.~\cite{FisherMaNickel_PRL_1972}.
This model was particularly convenient since the exponent $\eta$ of its Wilson-Fisher FP is exactly known, thus saving 
the numerical effort to determine it numerically from the simulation data. In addition, detailed analytical calculations and quite precise 
Monte Carlo data were available for comparison \cite{Luijten_PhD_1997}. Using our FMC implementation of MSRG, we were indeed able to observe the $b$-dependence of $\bKeff^*(b)$ and its associated exponents.
However, contrary to our initial expectations, it turned out that the best accuracy was not obtained in the large $b$ limit.
Instead, for varying $b$ the FP $b\mapsto \bKeff^*(b)$ moves along a ``trajectory'' in the plane $\Keff$ that exhibits a turning point at a certain shell thickness $b^*$
that was actually found to be rather close but distinct from $b=1$, and for this distinguished value $b^*$ we observed that the values of the critical exponents $\nu$ and $\om$ 
were in excellent agreement with the benchmark results derived in Ref.~\cite{Luijten_PhD_1997}. 
A systematic study for different system sizes revealed the surprising discovery that $1<b^*<\infty$ is not a finite size effect.
Nevertheless, we speculate that this peculiar finding is highly specific to the model of  Fisher, Ma and Nickel, and we still expect that usually $b^*\to1$ or $b^*\to\infty$ will instead be found     
in other systems.
The rest of the paper will therefore be devoted to the application of our ideas to a real-world system, whose critical properties are still
an active area of research: the elastic behavior of crystalline membranes.

\section{FMC Implementation of MSRG for crystalline membranes}

As explained in detail in Refs.~\cite{NelsonPiranWeinberg_Membranes_1988,Safran_STSIM_2003}, the flat phase of a crystalline membrane is conveniently described in the so-called Monge parametrization, 
which amounts to specifying a scalar ``height'' function $f(x)$ that measures the out-of-plane deformations of the membrane with respect to a two-dimensional reference plane,
which we take to be of size $L\times L$ with periodic boundary conditions understood. 
The long-wavelength physics of the system is captured by the Fourier modes 
\begin{eqnarray}
\tilde f(\bq)=\theta(\La-|\bq|)\int d^2x f(\bx)e^{-i\bq\bx} \,,
\end{eqnarray}
where the Heaviside step function is used to impose a cutoff $\La$ in the space of wavevectors.
Formally embedding the vectors $\bq,\bQ$ in $\mathbb{R}^3$ and abbreviating $\wh\bQ=\bQ/|\bQ|$, we define 
\begin{eqnarray}
\tilde\calF(\bQ)=\int\frac{d^2q}{(2\pi)^2} \left(\wh\bQ\times\bq\right)^2 \tilde f(\bq)\tilde f(\bQ-\bq) \,.
\label{eqn:mmsksmkwswkwsmkwsmkwskwsxjsxnsjxnsjsxxm2}
\end{eqnarray}
In terms of this generalized convolution,  the effective Hamiltonian
that describes the universal properties of the flat phase at long wavelengths is the given by
\begin{eqnarray}
\calH^\La[f]=\!\frac{\kappa}{2}\int \!\frac{d^2q}{(2\pi)^2} q^4 |\tilde f(\bq)|^2
+\frac{K}{8}\!\int\!\frac{d^2Q}{(2\pi)^2}|\tilde\calF(\bQ)|^2 \,.
\label{eqn:hwchwcvwejhcvjgwcewhhewwhewvwvwvwvwgvg}
\end{eqnarray}
Its first contribution, the bending energy, is represented by a local dispersion term as found in a standard LG model, except that 
the usual gradient term $(\nabla f)^2$ is replaced by a Laplacian $(\Delta f)^2$. In addition, while its second contribution also involves
four powers of $f$, the non-local characteristic of the generalized convolution (\ref{eqn:mmsksmkwswkwsmkwsmkwskwsxjsxnsjxnsjsxxm2})
hints at physics that is quite different from that of the standard LG model. 
As discussed in Refs.~\cite{NelsonPiranWeinberg_Membranes_1988,Katsnelson_Graphene_2012}, this non-locality encodes an effective long-range
anharmonic self-interaction of the out-of-plane deformations $f$ mediated by in-plane phonons
that had been integrated out in the calculation steps leading to (\ref{eqn:hwchwcvwejhcvjgwcewhhewwhewvwvwvwvwgvg}).
$\calH[f]$ involves only two coupling constants, namely the bending stiffness $\kappa$ and
the the effective 2d Young modulus $K=4\mu(\mu+\la)/(2\mu+\la)$ composed from the in-plane Lame constants $\la$ and $\mu$ of the membrane.
Implicit in all these constants as well as in the formulas (\ref{eqn:hwchwcvwejhcvjgwcewhhewwhewvwvwvwvwgvg} but suppressed in our present notation 
is a dependence on the cutoff $\La$.

To implement our FMC algorithm, we replace the membrane reference plane by a square $L\times L$ lattice with $N=L^2$ sites and lattice constant $a=1$, and keep the imposed periodic boundary conditions.
Assuming without loss of generality $L$ to be even, we may parametrize
wavevectors inside the full first Brillouin zone of this lattice by $q_i=2\pi m_i/L$, $m_i=-L/2+1,\dots,0,\dots,L/2$, and the above integrals over the Brillouin zone are replaced by finite sums. 
In view of the rectangular structure of the underlying lattice, it is natural to replace spherical cutoffs $\La$ that are convenient in analytic continuum calculations
by a more suitable cubic version. Parametrized by an integer $l$, in our simulation a cutoff $\La=2\pi l/L$  is applied to each separate wavevector component,
and in order to avoid problems with ``umklapp'' terms and minimize effects of lattice anisotropy, it is recommended to choose $l\ll L/2$. 
To implement the coarse graining step in FMC, we furthermore choose an inner cutoff $\La'=2\pi l'/L$ with $0<l'<l$. The shell thickness parameter 
is then given by $b=\La/\La'=l/l'$.

The discrete Fourier transform convention 
\begin{eqnarray}
\tilde f(\bq)=\entwederoder{\sum_{\bx}f(\bx)e^{-i\bq\bx}}{|q_i|<\La}{0}{\text{else}}  
\end{eqnarray}
with inversion
\begin{eqnarray}
f(\bx)=\frac{1}{N}\sum_{|q_i|<\La}\tilde f(\bq)e^{i\bq\bx}  \,,
\end{eqnarray}
in which the Fourier amplitudes are extensive quantities, may look somewhat asymmetric, it proves to be convenient in comparing
discrete to continuous formulas.
Since the membrane's elastic free energy does not depend on the average distance of the membrane to the Monge reference plane but merely on variations of its height,
only derivatives of $f$ enter in the in the continuum formulation (\ref{eqn:hwchwcvwejhcvjgwcewhhewwhewvwvwvwvwgvg}). Therefore
we can further assume without loss of generality that $\tilde f(\bnull)=0$.
In terms of these discrete amplitudes, the above formulas (\ref{eqn:mmsksmkwswkwsmkwsmkwskwsxjsxnsjxnsjsxxm2}) and (\ref{eqn:hwchwcvwejhcvjgwcewhhewwhewvwvwvwvwgvg}) are replaced by 
\begin{eqnarray}
\tilde\calF(\bQ)=\sum_{\bq}\left(\wh\bQ\times\bq\right)^2 \tilde f(\bq)\tilde f(\bQ-\bq) 
\end{eqnarray}
and
\begin{eqnarray}
\calH^\La[f]=\frac{\kappa_N}{2} \sum_{\bq\ne\bnull} q^4 |\tilde f(\bq)|^2
+\frac{K_N}{8}\sum_{\bQ\ne\bnull}|\tilde\calF(\bQ)|^2\,,
\label{eqn:xqsjkxjkxnjxnjqxnjqnjqnjqnqwjnqw}
\end{eqnarray}
where
\begin{eqnarray}
\kappa_N=\frac{\kappa}{N},\quad K_N=\frac{K}{N^3}  \,.
\end{eqnarray}

In view of the extensive discussions already available in the literature (cf.~Refs.~\onlinecite{TroesterDellago_F354_2007,TroesterCPC179_2008,Troester_CSSCMP_2008,Troester_PP53_496_2014})
and the detailed layout of the specific implementation for the case of crystalline membranes presented in the companion paper \cite{Troester_PRB_87_104112_2013}, we would like to keep
the description of the basic Fourier Monte Carlo algorithm and its general properties at a minimum in the present paper.
However, it turns out that setting up the coarse graining step of MSRG for a crystalline membrane requires to define different MC moves for slow and fast modes of
the so-called ``tracer'' configurations to be defined below.  
In the standard cubic FMC scheme the momentum shell corresponding to a prescribed pair of cutoffs $\La'<\La$ is, of course,
defined as the set of wave vectors with components $p_i,i=1,2$ subject to the constraints
$|p_i|\le\La$ and  $\max_{i}|p_i|>\La'$. 
MC move of fast modes $\tilde f_>(\bq)$ are performed by picking a random wave vector $\bp$ from this shell, 
choosing a random complex number inside a circle $|\eps|<\rho$ of radius $\rho$ around $0$  in the complex plane, and considering the shift 
\begin{eqnarray}
\tilde f(\bq)\to \tilde f(\bq)+\eps\delta_{\bq,\bp}+\eps^*\delta_{\bq,-\bp}\,. 
\label{eqn:ncjdcnjxcnlscnckck}
\end{eqnarray}
Taking advantage of the special convoluted structure (\ref{eqn:mmsksmkwswkwsmkwsmkwskwsxjsxnsjxnsjsxxm2}) of the anharmonic term
appearing in (\ref{eqn:hwchwcvwejhcvjgwcewhhewwhewvwvwvwvwgvg}), it is then possible to calculate the resulting change in energy 
in an efficient way, as is explained in detail in Ref.~\cite{Troester_PRB_87_104112_2013}.

Integrating out these fast modes by means of an FMC simulation should then produce a coarse-grained Hamiltonian of general structure
\begin{widetext}
\begin{eqnarray}
\tilde\calH^{\La/b}[f]=\frac{1}{2}\sum_{\bq}\left[\tilde\kappa_N q^{4}+\dots\right]|\tilde f(\bq)|^2
+\frac{1}{8}\sum_{\bQ}\left[\tilde K_N +\dots\right]|\tilde\calF(\bQ)|^2+O(f^6)
\label{eqn:DHDBQDBQBBD}
\end{eqnarray}
\end{widetext}
with
\begin{eqnarray}
\tilde\calF(\bQ)=\sum_{\bq} \left(\wh\bQ\times\bq\right)^2 \tilde f(\bq)\tilde f(\bQ-\bq)\,, 
\label{eqn:DHDBQDBQBBD2}
\end{eqnarray}
from we wish to extract the two CG relations $\kappa_N\mapsto\tilde\kappa_N$ and $K_N\to\tilde K_N$,
i.e.~$\kappa\mapsto\tilde\kappa$ and $K\to\tilde K$.
For this purpose, we determine the value of the CG Hamiltonian (\ref{eqn:DHDBQDBQBBD}) by restricting the MC sampling to 
certain ``tracer configurations'' defined by a particularly simple and convenient choice of their slow mode parts. 
In terms of simplicity, our preferred type of such a tracer configuration would certainly be that of an isolated ``dumbbell''
of just two slow modes with a common uniform real-valued amplitude at the fixed wave vector $\pm\bk$.
This dumbbell is surrounded by the shell of nonzero fast modes, but all other slow modes are put to zero. In formal terms,
the slow parts of such dumbbell tracer configurations $\tilde f^{(\bk)}(\bq)$ defined with respect to $\pm\bk$ are restricted to be of type 
\begin{eqnarray}
\tilde f_{<}^{(\bk)}(\bq)\equiv f_d(\delta_{\bq-\bk}+\delta_{\bq+\bk}),\ f_d\in\mathbb{R} \,.
\label{eqn:xxmxmkmkmkmkqmkmdkqmkqwmqwkmqwk} 
\end{eqnarray}
As explained in detail in Refs.~\cite{Troester_PRB76_2007,Troester_CSSCMP_2008}, 
for this class of tracer configurations one now performs a multicanonical type of simulation of e.g.~the Wang-Landau type,
in which the probability distribution $P(f_d)$ of the "reaction coordinate" $f_d$ in the ``bath'' of fast modes is calculated. A polynomial fit of
$-\ln P(f_d)$ then yields a set harmonic and lowest order anharmonic coefficients $a_2(\bk),a_4(\bk),\dots$ for each chosen wave vector $\bk$. 
Comparison of these coefficients with the general $\bk$-dependent structure of the bare effective Hamiltonian then allows to determine
a ``new'' set of bare parameters. In other words, one obtains all the information required for completing the coarse graining step of the MSRG prescription.    
   
For LG type of models with a local anharmonic energy contribution, this class of tracer configurations allows to determine
the flow of coupling parameters. Unfortunately, however, for our present problem the dumbbell class (\ref{eqn:xxmxmkmkmkmkqmkmdkqmkqwmqwkmqwk}) is
insufficient to capture the flow of the anharmonic part of the bare Hamiltonian. In fact, in the formula
\begin{eqnarray}
\tilde\calF_<^{(\bk)}(\bQ)= (\hat\bQ\times\bk)^2 \left[\delta_{\bQ-2\bk}+2\delta_{\bQ}+\delta_{\bQ+2\bk}\right]f_d^2
\end{eqnarray}
that results for amplitudes $\tilde\calF(\bQ)$ built exclusively from the slow part $\tilde f^{(\bk)}_<(\bq)= f_d(\delta_{\bq-\bk}+\delta_{\bq+\bk})$,
the vector $\bQ$ is constrained to be either zero (which is forbidden) or parallel to $\bk$, in which case the leading cross product vanishes,
i.e.~$\tilde\calF_<^{(\bk)}(\bQ)\equiv 0$.
To overcome this difficulty, we instead consider ``cross'' tracer configuration with slow parts of type
 \begin{eqnarray}
\tilde f^{\bk}_{<}(\bq)=f_c\left(\delta_{\bq-\bk}+\delta_{\bq-\bk^\perp}+\delta_{\bq+\bk} +\delta_{\bq+\bk^\perp}\right)\,,
\label{eqn:xxjkxnjknxjnxjnxjnjknjjknjkn} 
\end{eqnarray}
where $f_c\in\mathbb{R}$ and $|\bk|=|\bk^\perp|,\, \bk\cdot\bk^\perp=0$, 
which map out a symmetric ``cross'' spanned by two orthogonal vectors $\bk$ and $\bk^{\perp}$ of equal length around $\bnull$ with one common 
real-valued amplitude, all remaining slow modes being 
silenced to zero.
For this class of tracer  configurations, a lengthy but elementary calculation yields
\begin{widetext}
\begin{eqnarray}
\tilde\calF^{\bk}(\bQ)
=2(\hat\bQ\times\bk)^2f_c^2\left[\delta_{\bQ-\bk-\bk^\perp}+\delta_{\bQ+\bk-\bk^\perp}+\delta_{\bQ-\bk+\bk^\perp} +\delta_{\bQ+\bk^\perp+\bk}\right]\,.
\end{eqnarray}
\end{widetext}
In particular, if the arms of the cross are chosen to point along the directions $\bk=(k,0),\ \bk^\perp=(0,k)$
of the Cartesian axes, we have
\begin{eqnarray}
\left(\hat\bQ\times\bk\right)^2\delta_{\bQ\pm\bk\pm\bk^\perp}=\frac{k^2}{2}\delta_{\bQ- (\pm k,\pm k)}\,,
\label{eqn:kcmkcmkmckmdkmdmkdmqdmq}
\end{eqnarray}
and the above equations simplifies to
\begin{eqnarray}
\tilde\calF^{\bk}(\bQ)=k^2f_c^2\cdot \entwederoder{1}{\bQ=(\pm k,\pm k)}{0}{\text{else}}\,.
\label{eqn:xjxjxnjnjnxxnxnxnn}
\end{eqnarray}
Using this result, we calculate the total energy contribution of a cross configuration without fast modes as 
\begin{eqnarray}
E^{\bk}(f_c)=2\kappa_Nk^4f_c^2+\frac{K_N}{2}k^4f_c^4\,.
\label{eqn:xx3rfrbfrfgrffbrgfburfb3hrbf3uhbf3uh3u}
\end{eqnarray}
From the MC point of view (\ref{eqn:xxjkxnjknxjnxjnxjnjknjjknjkn}) imposes an extra constraint on the allowed phase space in addition to
the reality condition $\tilde f(\bk)=\tilde f^*(-\bk)$ for the fast modes during the sampling. 
We thus need to calculate the effect of a variation 
\begin{eqnarray}
\delta f(\bq)=r\left(\delta_{\bq-\bk}+\delta_{\bq-\bk^\perp}+\delta_{\bq+\bk} +\delta_{\bq+\bk^\perp}\right)
\label{eqn:hqhbhqdqdqdqdqqqdq}  
\end{eqnarray}
of the cross configuration by the real number $r$ on the total energy.  
For the harmonic contribution, it is easy to see that
\begin{eqnarray}
\delta E_{\text{harm}}=4\kappa_Nk^4(rf_c+r^2/2)
\end{eqnarray}
It remains to calculate the change of the anharmonic contribution to the energy under a MC move (\ref{eqn:hqhbhqdqdqdqdqqqdq}). In terms of the shift $\delta\tilde\calF(\bQ)$, for which   
a lengthy and tedious calculation yields
\begin{widetext}
\begin{eqnarray}
\delta\tilde\calF(\bQ)=
2r(\wh\bQ\times\bk)^2\left[\tilde f(\bQ-\bk)+\tilde f(\bQ+\bk)+r\delta_{\bQ-\bk-\bk^\perp}\right]+(\bk\leftrightarrow\bk^\perp)\,.
\label{eqn:cnjqbnqejnjdibdebdiqdbiqedbibdidbq}
\end{eqnarray}
\end{widetext}
This last missing piece of information is readily obtained from the general variation formula
\begin{eqnarray}
\delta E_{\text{anharm}}=\frac{K_N}{8}\sum_{\bQ\ne\bnull}\left[2\tilde\calF(\bQ)\delta\tilde\calF(-\bQ)+|\delta\tilde\calF(\bQ)|^2\right]  
\end{eqnarray}
valid for all types of FMC moves.

Recently \cite{Troester_PRB_87_104112_2013} we have introduced a new variant of FMC that is able to efficiently suppress critical slowing down
i.e.~exponential growth of integrated autocorrelation times in critical or nearly critical systems. 
This is achieved by iteratively optimizing the MC acceptance rates of individual Fourier amplitudes during the start-up phase of the simulation 
for each wave vector separately, aiming at acceptance rates between $30\%-40\%$ for each amplitude.
In the present simulations, such an optimization was, of course, also implemented.

In what follows, we shall assume that without loss of generality $\kappa=1$,
such that only a dependence on the anharmonic coupling parameter $K$ remains.
The coarse graining procedure outlined so far produces a shift 
\begin{eqnarray}
\bK:=\left(1\atop K\right)\mapsto\left(\tilde\kappa\atop\tilde K\right)=:\tilde \bK\,.
\end{eqnarray}
According to (\ref{eqn:cnhcbjhcbhcbbqbqab}), rescaling of lengths and further ``wave function'' renormalization then leads to
\begin{eqnarray}
\tilde \bK\mapsto\left(b^{-\eta(K)}\tilde\kappa\atop b^{2-2\eta(K)}\tilde K\right)=:\left(\kappa'\atop K'\right)=:\bK'
\end{eqnarray}
since $[\kappa]=0$ and $[K]=2$.
A concrete RG flow $\bK\to\bK'$ is only defined after specifying the function $\eta(K)$. 
Imposing invariance $\kappa'\equiv 1$ of the harmonic dispersion term gives
\begin{eqnarray}
\eta(K)=\frac{\ln \tilde\kappa(K)}{\ln b}\,,
\label{eqn:cklcccmcklcmdklmcdklcmdklcdkappa}
\end{eqnarray}
where we explicitly indicate the dependence of $\tilde\kappa$ on the parameter $K$.
On the other hand, an invariance condition $K'\equiv K$ would implicitly define a function $\eta_K(K)$ by 
\begin{eqnarray}
\eta_K(K)\equiv 1+\frac{\ln\frac{\tilde K(K)}{K}}{2\ln b} \,.
\label{eqn:cklcccmcklcmdklmcdklcmdklcdK}
\end{eqnarray}
At a FP $K'\equiv K\equiv K^*$, the common value 
\begin{eqnarray}
\eta(K^*)=\eta_K(K^*)\equiv \eta  
\label{eqn:cklcccmcklcmdklmcdklcmdklcdequality}
\end{eqnarray}
of the two functions $\eta(K)$ and $\eta_K(K)$ is the critical exponent $\eta$.
Thus $K^*$ can be numerically determined as the location of the common intersection point of these functions plotted against $K$.

Since the RG transform $K\mapsto K'(K)$ is designed to be analytic, we can linearize it around the FP value $K^*$, such that
\begin{eqnarray}
K'(K^*+\delta K)\approx K^*+\calM\cdot\delta K\,,
\label{eqn:ejednendejee} 
\end{eqnarray}
where
\begin{eqnarray}
\calM:=\left.\frac{dK'(K)}{dK}\right|_{K=K^*}  
\end{eqnarray}
denotes the slope of the function $K\mapsto K'(K^*)-K^*$ at its zero $K=K^*$, which can readily be assessed in our simulations. 
For a nontrivial infrared attractive FP we expect that $K$ is irrelevant and thus $|\calM|<1$.
If $K'(K)$ is not to oscillate back and forth around this FP during successive RG iterations, we should also expect $\calM>0$.
The corresponding Wegner \cite{Wegner_PRB5_4529_1972} exponent $\om$ is then defined through $\calM\equiv b^{-\om}$, i.e. 
\begin{eqnarray}
\om=-\frac{\ln\calM}{\ln b} \,.
\label{eqn:djdnejdnjendjnejnjednwednjqwnd}
\end{eqnarray}

\section{Numerical Evaluation Strategy}

The membrane systems studied in our simulations may be parametrized by a triple of integers $(L,l,l')$, such that $\La=2\pi l/L,\La'=2\pi l'/L,b=l/l'$. 
Each of the integers $j=1,\dots,l'$ then defines a value
\begin{eqnarray}
k_j=2\pi j/L 
\label{eqn:chkdwcbcbdbckbcwbcljw}
\end{eqnarray}
for a cross (\ref{eqn:xxjkxnjknxjnxjnxjnjknjjknjkn}) with arms 
\begin{eqnarray}
\bk_j=(\pm k_j,0),\qquad \bk_j^\perp=(0,\pm k_j)
\label{eqn:chkdwcbcbdbckbcwbcljw1}
\end{eqnarray} 
that hosts a tracer configuration with real-valued amplitude $f_c$. 
In principle, each choice of inner cutoff parameters $1\le l'<l$ gives rise to $k$-values $k_1,\dots,k_{l'}$.
In Ref.~\cite{Troester_PRB_87_104112_2013} we have observed strong finite size irregularities for the correlation function $\tilde G(\bk)=\langle \tilde f(\bk)\tilde f(-\bk)\rangle$
at the smallest accessible nonzero wave vectors, so that it is recommended to exclude the $k$-value $k_1$ from the following fits. 
Since one needs a minimum of approx.~$5$ to $6$ values to determine dispersions with sufficient statistical reliability (see below), 
we are confined to lower cutoffs of about $l'\ge 6$, which in turn puts an approximate lower limit of $1/b\ge 6/l$ 
on the $b$-values accessible in our simulations.
On the other hand, due to the discreteness of the Brillouin zones of our finite systems, the closest accessible value of $1/b$ below its ultimate limit $1.0$ is
$1/b=(l-1)/l$. 

\begin{figure}[tbp]
\centering
\includegraphics[scale=0.7]{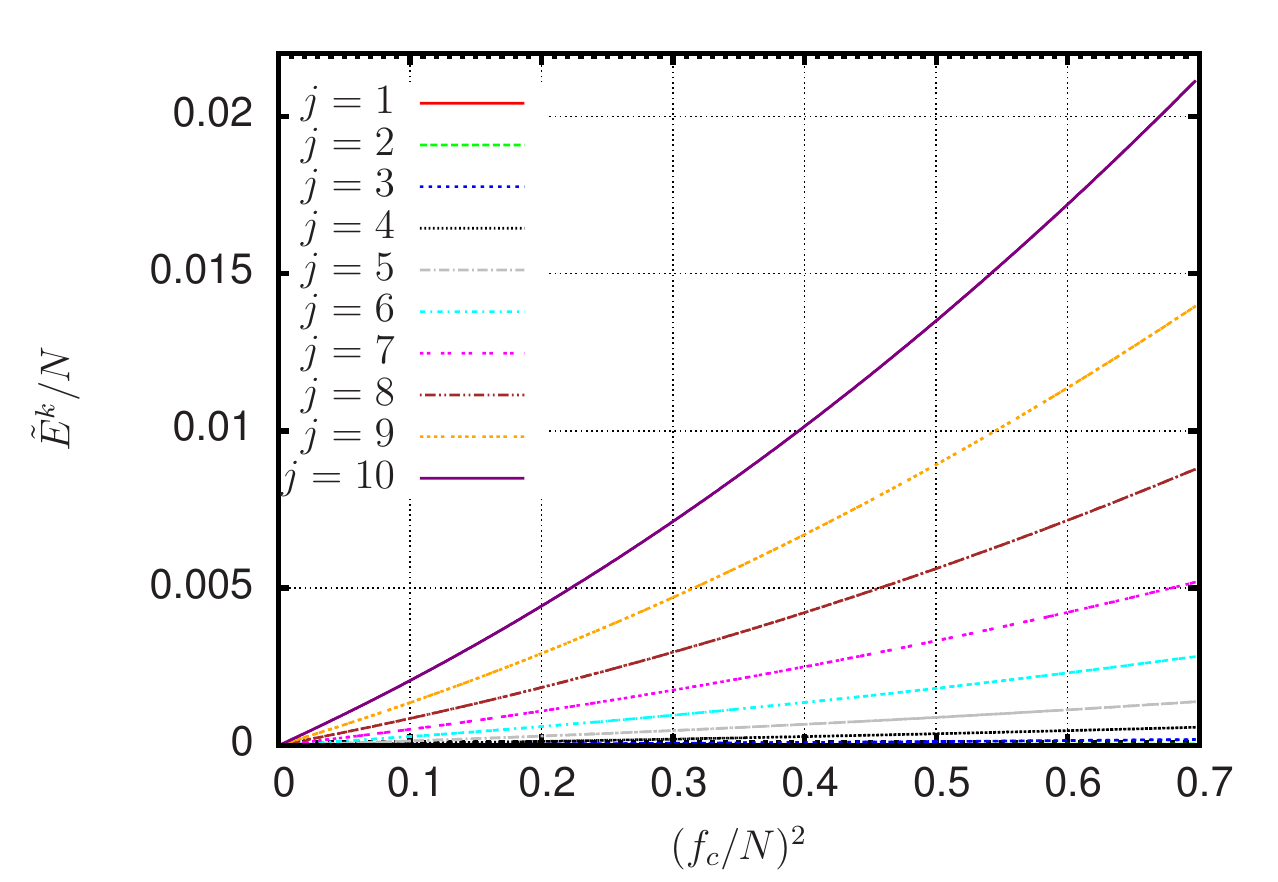}
\caption{Raw simulation data for systems of size $L = 240$
with outer cutoff $l = 24$, inner cutoff $l' = 10$ and 2d Young modulus
$K = 9.16$. The index $j = 1,\dots,10$ labels the underlying tracer
configurations of type (\ref{eqn:xxjkxnjknxjnxjnxjnjknjjknjkn}) with stars (\ref{eqn:chkdwcbcbdbckbcwbcljw1}) parametrized by
the wave numbers $k_j$ as defined in Eqn. (\ref{eqn:chkdwcbcbdbckbcwbcljw}).}
\label{fig:freenergy}
\end{figure}

Given such a system, we determine the unnormalized probability distribution $P^{\bk}(f_c)$
within a certain interval $-f_{\mathrm{max}}\le f_c\le f_{\mathrm{max}}$ by FMC, and thus the dimensionless coarse-grained free energy   
\begin{eqnarray}
\tilde E^{\bk}(f_c):=-\ln \frac{P^{\bk}(f_c)}{P^{\bk}(0)}\,.
\end{eqnarray}
To reliably separate the contributions proportional to $f_c^2$ from those proportional to $f_c^4$ in a comparison of 
(\ref{eqn:xx3rfrbfrfgrffbrgfburfb3hrbf3uhbf3uh3u}) to these data requires to choose a suitable value for $f_{\mathrm{max}}$.  
To estimate this value, we analyze the bare energy expression (\ref{eqn:xx3rfrbfrfgrffbrgfburfb3hrbf3uhbf3uh3u}), assuming that
its coarse-grained counterpart will not differ by orders of magnitude from it.
After fixing $\kappa\equiv 1$, our only remaining parameter is the value of the remaining bare coupling parameter $K$. 
Since both the harmonic as well as the anharmonic contribution to (\ref{eqn:xx3rfrbfrfgrffbrgfburfb3hrbf3uhbf3uh3u})
are proportional to $k^4$, it seems reasonable to choose a common value $f_{\mathrm{max}}$ for the amplitude at which we expect 
to see a factor of $\la$  between the bare total energy $E^{(c)}$ and its purely harmonic part $E^{(h)}\Big|_{K_N=0}$ 
uniformly for all $k$. Numerically, $f_{\mathrm{max}}$ is determined from the equation
\begin{eqnarray}
2\kappa_Nf_{\mathrm{max}}^2+\frac{K_N}{2}f_{\mathrm{max}}^4\equiv \la\cdot 2\kappa_Nf_{\mathrm{max}}^2 
\end{eqnarray}
i.e.
\begin{eqnarray}
f_{\mathrm{max}}=2N\sqrt{\frac{(\la-1)\kappa}{K}}
\end{eqnarray}
If $\la$ is chosen too small or too large, it becomes numerically hard to reliably separate harmonic and anharmonic contributions by a least squares fit.
Moreover, the free energy range that needs to be determined in the simulations increases with growing $\la$.
After performing various numerical tests, we settled for a common factor of $\la=2.6$ which was used in all subsequent simulations.  
To actually explore the potential shape in this region, a successful simulation approach needs to overcome 
potentially large free energy differences. After monitoring the convergence and tunneling 
properties of several variants of the family of multicanonical
algorithms, the $1/t$ variant \cite{Belardinelli_JCP127_2007,Belardinelli_PRE_2008} of the Wang-Landau algorithm \cite{WangLandau_PRL_2001,Landau_BrazJPhys34_354_2004} emerged as a robust and reliable choice. 
Each single simulation was performed with
an order of magnitude of $200$ tunneling events between $f_c=0$ and $f_c=f_{\mathrm{max}}$ at the $1/t$ stage, which took about $10^6$ single MC sweeps per simulation,
and for each value of $K$ that we want to inspect, $l'$ such simulations are needed, one for every $k$-value out of the set $\{k_j:j=1,2,\dots,l'\}$.
An example of the raw data obtained by such simulations may be inspected in Fig.~\ref{fig:freenergy}.

The numerical procedure to determine RG recursion relations requires therefore a number of nested least squares fits.
On the one hand, the function $\tilde E^{\bk}(f_c)$ obtained from a simulation of the above type will, of course, be contaminated
by small contributions of powers higher than $f_c^4$, i.e.~it will not exactly resemble the simple structure of the bare effective Hamiltonian (\ref{eqn:hwchwcvwejhcvjgwcewhhewwhewvwvwvwvwgvg}).
This well-known behavior of generating ``new'' couplings beyond those present in the original bare Hamiltonian, which is inherent to the RG, can only be dealt with
by fitting to a more general ansatz of type
\begin{eqnarray}
\tilde E^{k}(f_c)\equiv a(k)f_c^2+b(k)f_c^4+ c(k)f_c^6+ d(k)f_c^8
\label{eqn:bhdbhdbhdbqwhb}
\end{eqnarray}
and discarding all coefficients except $a(k)$ and $b(k)$ in the analysis that follows. 
In a second level of fitting, the resulting collection of coefficients $\{a(\bk),b(\bk)\}$ is in turn fitted to functions of structure  
\begin{eqnarray}
a_{\text{fit}}(k)&\equiv&2\tilde\kappa_Nk^4     +a_6k^{6}+a_{8}k^{8}+\dots+ a_{n_{\mathrm{max}}}k^{n_{\mathrm{max}}}\,,\\
b_{\text{fit}}(k)&\equiv&\frac{\tilde K_N}{2}k^4+b_6k^{6}+b_{8}k^{8}+\dots+ b_{n_{\mathrm{max}}}k^{n_{\mathrm{max}}}\,,
\label{eqn:snjnsjwsnjwsnjnjwnjw}
\end{eqnarray}
from which the coarse-grained values $\tilde\kappa_N,\tilde K_N$ are extracted, while all higher order fit parameters, which correspond to other higher order couplings
presumably generated by the coarse graining operation, are ignored once again.

In practice these fits are not as straightforward to do as it may seem. Our numerical tests did show that for the first level of fitting,
using the truncated eight order polynomial (\ref{eqn:bhdbhdbhdbqwhb}) as a fit function  order gave numerically convincing results for all considered parameter ranges. 
However, it turned out to be very hard to decide in advance 
how to choose the maximum power $n_{\mathrm{max}}$ of $k$ kept in the definition of the fit functions (\ref{eqn:snjnsjwsnjwsnjnjwnjw}) in the second level fitting that
aims at extracting the lowest order $\bk$-dependence of the functions $a(\bk)$ and $b(\bk)$. Truncating at too low orders may
yield a certain trade-off among the resulting fit parameters and thus adulterate the results. On the other hand, remember that we are necessarily working
in a finite system with a discrete Brillouin zone, the minimum spacing between $\bk$-vector components given by $2\pi/L$, and so only few data points may be available for 
low inner cutoff parameter $l'$. Specifically,
for systems with small values of, say, $l'\le 6$, a high order polynomial fit will not produce meaningful results, since
too few unintegrated modes with small $\bk$-vectors parallel to a chosen direction are at our disposal.

Worse, the importance of higher order terms in the expansion was observed to strongly vary with the particular value of $K$ chosen. 
The ``optimal'' truncation order $n_{\mathrm{max}}$ of the polynomials may thus even depend on $K$, which makes it very difficult to evaluate the large mass of data generated 
in our simulations in this way.
Worst, it may be difficult to figure out possible ``forbidden'' powers in the sought-after expansion. For instance, we have explicitly checked numerically that 
no ``surface tension'' contribution  $\propto k^2$ to $a(k)$ is generated from (\ref{eqn:xqsjkxjkxnjxnjqxnjqnjqnjqnqwjnqw}) by the coarse graining operation, 
which justifies a posteriori the use of (\ref{eqn:xqsjkxjkxnjxnjqxnjqnjqnjqnqwjnqw})
as our basic model Hamiltonian. Theoretically, this absence can be contributed to the presence of a Ward identity (see e.g.~the cancellation of $k^2$-contributions in the sum
of contributions to Eqs.~(30a-d) of Ref.~\cite{Hasselmann_PRE83_031137_2011}). However, it is beyond the scope of the present work to compute all similar constraints  
on higher order $k$-dependent expansion coefficients imposed by Ward identities.

To summarize the above observations, we need to fit data obtained for the collection of $\bk$-vectors 
with a function of which only the lowest expansion power is known with certainty. This problem may look hopeless or at least somewhat ill-defined at first sight.
Not being aware of any pre-assembled approach published in the literature on numerical mathematics, we had to come up with our own custom solution.
As is explained in more detail in the Appendix,  
the basic philosophy of our approach is not to focus on the unknown higher order contributions to the fit functions, but rather to determine the extent of validity of the lowest order 
approximation to the underlying data set and reweighting the members of this data set accordingly.
Despite currently lacking a rigorous mathematical proof, this seems to work quite well in practice.

\begin{widetext}

\begin{figure}[tbp]
\centering
\includegraphics[scale=1.1]{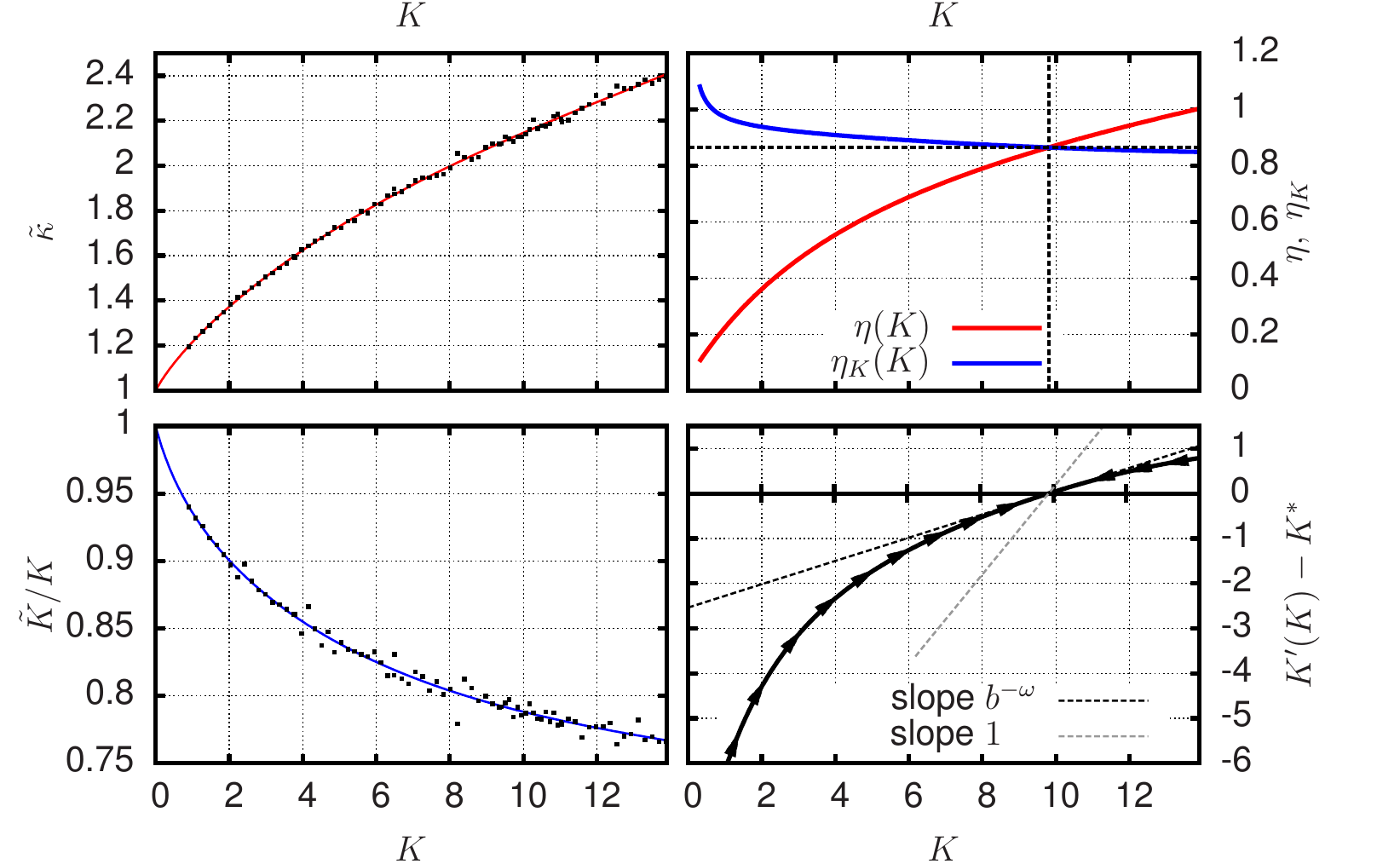}
\caption{Illustration of our numerical procedure for parameters $L=240$, $l=24$ and $l'=10$.
Left column plots:  fits of data for $\tilde\kappa(K)$ and $\tilde K(K)/K$ to the ansatz (\ref{eqn:xxxxksxkmxkmxkmxkmxkxmkmxmxk}). Each data point shown is derived from fits of the functions $a(k),b(k)$ based on
up to $l'$ different simulation data of the type shown in Fig.~\ref{fig:freenergy}.
Note the excellent compliance of the numerical data with the necessary conditions  $\lim_{K\to 0}\tilde\kappa(K)
=\lim_{K\to 0}\tilde K(K)/K=1$. 
Right column, upper plot: illustration of the numerical solution of the equations $\eta(K)\equiv \eta_K(K)$ as defined in Eqs.~(\ref{eqn:cklcccmcklcmdklmcdklcmdklcdkappa}), (\ref{eqn:cklcccmcklcmdklmcdklcmdklcdK}).
Right column, lower plot: determination of the slope of function $K'(K)-K^*$ at $K=K^*$. The direction of the RG flow is schematically indicated.
In this particular example, the intersection point determined by Eqn.~(\ref{eqn:cklcccmcklcmdklmcdklcmdklcdequality}) is located at $K^*=9.80127$, 
and we derive exponents $\eta=0.86507$ and $\om=1.54456$, respectively. 
}
\label{fig:etaschnitt}
\end{figure}

\begin{figure}[tbp]
\centering
\includegraphics[scale=1.1]{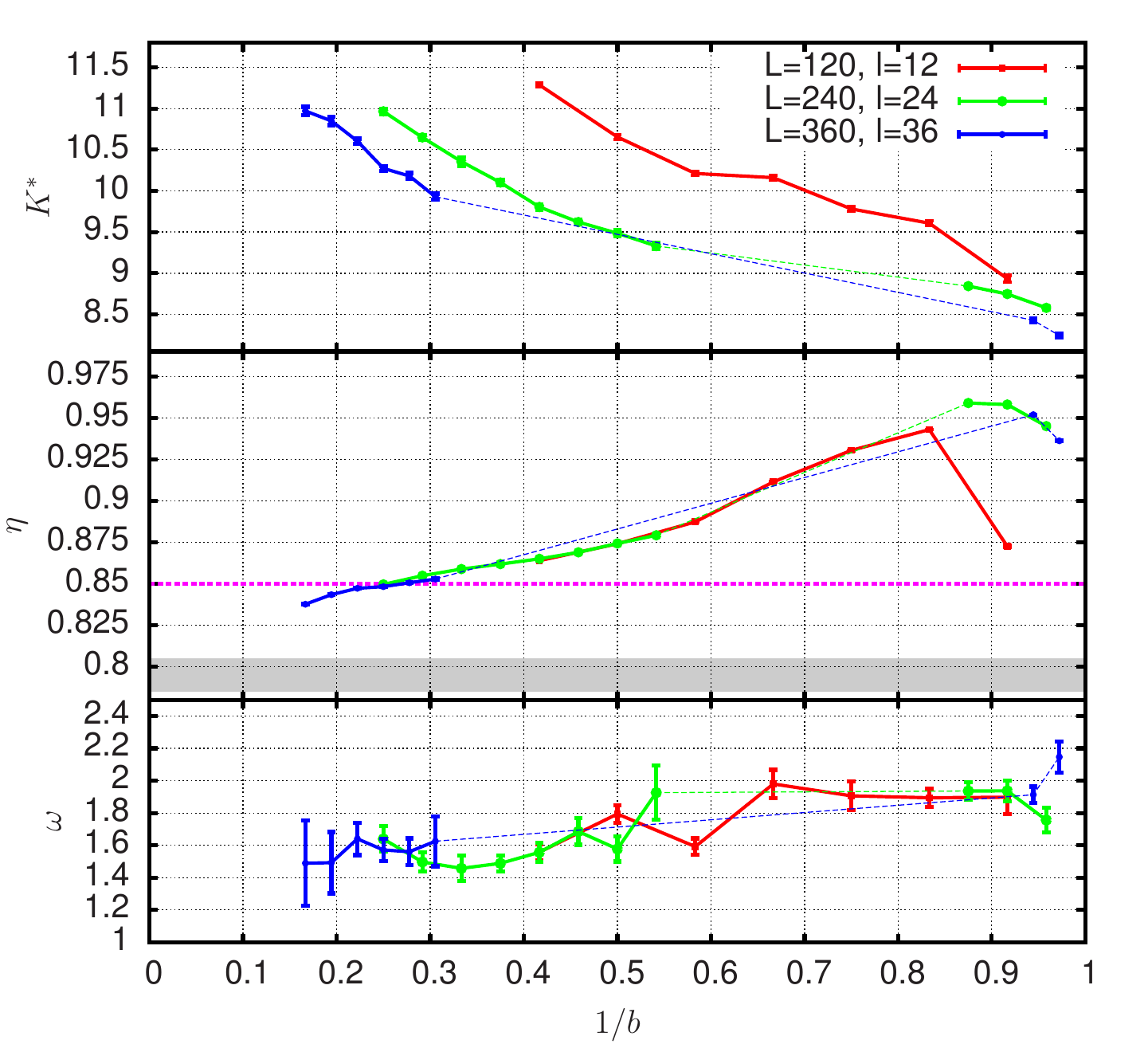}
\caption{Numerical results obtained for systems of sizes $L=120,240,360$ with outer cutoff parameters $l=12,24,36$, respectively. 
Upper plot: $b$-dependent location $K^*(b)$ of fixed point value of parameter $K$. 
Middle plot: $b$-dependent value $\eta(b)$ of exponent $\eta$. Lower plot: $b$-dependent value $\om(b)$
of exponent $\om$.}
\label{fig:ergvonb}
\end{figure} 
\end{widetext}

Putting $\kappa=1$ without loss of generality, the above procedure maps every ``bare'' coupling constants $K$ and shell thickness parameter $b$ into a pair
of coarse-grained coupling constants $(\tilde\kappa(K),\tilde K(K))$.
To determine the fixed point $K^*$ of the underlying RG transformation, it is thus necessary to perform
simulation for a large number of different $K$-values at each accessible $b$-value to extract the values for the fixed point coupling $K^*(b)$ and the exponents $\eta(b)$ and $\om(b)$ for various system sizes.
As explained above, for each such $K$-value this required $l'$ separate Wang-Landau type simulations (one for each $k_j$-value) to determine the underlying dispersion changes that
govern the computation of exponent $\eta$. For a numerical study it is advantageous to switch to description that is continuous in $K$, at the same time smoothing  the statistical noise contained in the resulting data.
To analyze the possible $K$-dependence of $\tilde\kappa(K)$ and $\tilde K(K)$, we make use of the simple fact that for  
$K\to 0$ evidently both $\tilde\kappa(K)\to \kappa=1$ and $\tilde K(K)/K\to 1$, such that any analytic fit of these functions must start out with value unity at $K=0$.
Otherwise, the only obvious requirement one may impose on a candidate fit function is that it should give a smooth and regular interpolation of the data. 
Numerical tests have shown that an ansatz of type
\begin{eqnarray}
f(K)= 1 + a \ln(1 + b^2 K) + c K + d K^2 + e K^3   
\label{eqn:xxxxksxkmxkmxkmxkmxkxmkmxmxk}
\end{eqnarray}
with five free parameters $a,\dots,e$ does a good job in this respect both for $\tilde \kappa(K)$ as well as for $\tilde K(K)/K$ (cf.~Fig.~\ref{fig:etaschnitt} for illustration).
With these analytic interpolations at hand, it is now an easy task to carry out the analysis 
outlined in Eqs.~(\ref{eqn:cklcccmcklcmdklmcdklcmdklcdkappa})-(\ref{eqn:djdnejdnjendjnejnjednwednjqwnd}).  

Error bars for the numerical quantities derived by these calculations are derived using a corresponding bootstrap analysis \cite{EfronTibshirani1994} based on
100 bootstrap samples drawn from the underlying set of considered coupling values $K$ for each particular choice of $(L,l,l')$.

\section{Results}

In the present work, we studied three systems with one common value of $\kappa=1$ and $\La=\pi/5$, with sizes 
defined by the parameters $(L,l)=(120,12),\,(240,24)$ and $(360,36)$, respectively.
Unfortunately, a complete scan through all available $b$-values was only possible for the smallest of these system due to the sheer amount of required computer resources. 
The resulting fixed point coupling value $K^*(b)$ for $L=120$, which is shown in Fig.~\ref{fig:ergvonb}, indicates a monotonous fall throughout the whole accessible range.
For $L=240$, the observed behavior is not in conflict with this hypothesis, and we would be very much surprised if the behavior for $L=360$ were fundamentally different.
Thus, it seems reasonable to assume that there is no critical maximum or minimum of $K^*(b)$ throughout the range $0<1/b^*<1$, which only leaves the possibilities $b^*=1$ or $b^*=\infty$.
Give a system defined by $(L,l,l')$, the one with $l'=l-1$ is of course one for which $1/b$ is closest to $1$. For our present purposes, thin momentum shells 
have some attractive features. For a thin shell the number of modes that need to be integrated out during the CG step is rather small, such that the simulations require less
cpu time than for thicker shells. At the same time, the large number of remaining unintegrated modes inside the shell should increase the numerical reliability of determining
the dispersions $a(k)$,$b(k)$. Unfortunately, however, there is a price to pay for this convenience. In fact, for a thin shell, all values $\tilde\kappa(K)$
and $\tilde K(K)/K$ were found to be extremely close to $1$ over the whole range of considered values of $K$, which represents a serious challenge to a numerical evaluation. 
For the larger two systems, the closest accessible estimates for $\eta$ in this limit are around $\eta\approx 0.93$, which is definitely out of range in comparison to 
all other published estimates. Although there may exist a common downward trend of $\eta(b)$ for $b\to1$, it is difficult to estimate the limiting behavior. 

Turning to the opposite limit $1/b\to0$, we observe a nice linear decrease of $\eta(b)$ with falling $1/b$, with all the data from various system sizes roughly
collapsing on a common same master curve, which indicates that for determining $\eta(b)$ finite size effects are small to negligible in this limit. 
Again, it is delicate to extrapolate the data to $1/b\to0$. If we assumed that the linear trend persists until $1/b\to0$, we would
arrive at a rough estimate of 
\begin{eqnarray}
\eta\approx 0.822\,,
\label{eqn:chbbchcchcbcbbw}
\end{eqnarray}
which is  quite satisfying, as it puts our present calculations roughly in the same ballpark as
those done analytically in the framework of the functional RG 
\cite{KownackiMouhanna_PRE79_040101_2009,Braghin_PRB82_035407_2010,Hasselmann_PRE83_031137_2011} where the estimate $\eta=0.85$ was derived.
Nevertheless, in view of the finite system size used, the imponderables of the above extrapolation $1/b\to 0$ and the fact that even in this limit the residual error due to the influence of
irrelevant couplings may only be minimized but not completely eliminated, one clearly should not expect (\ref{eqn:chbbchcchcbcbbw}) to be equally precise as the estimate 
$\eta = 0.795(10)$ of Ref.~\cite{Troester_PRB_87_104112_2013} derived from a systematic FSS analysis of the
membrane's mean squared displacement $\langle(\Delta f)^2\rangle$.

Extrapolation of $\om(b)$ to $b\to\infty$ is also delicate.
In fact, to an unprejudiced reader the data depicted in Fig.~\ref{fig:ergvonb} will be compatible with at least two scenarios:
\begin{itemize}
\item Linear extrapolation of $\om(b)$ to $b\to\infty$ produces a value of roughly 
$\om\equiv\lim_{b\to\infty}\om(b)\approx 4/3 \pm 0.3$.
\item The $b$-dependence of $\om(b)$ may just as well already have saturated for $b\to\infty$ at an asymptotically constant value, leading to any equally crude estimate of
$\om\approx 3/2 \pm 0.3$.
\end{itemize}
Even though these estimates may not seem to be extremely precise, they pave the way for a considerable further refinement of our previous FSS result as we show next.
To explain this in due detail, let us briefly recapitulate the approach followed in Ref.~\cite{Troester_PRB_87_104112_2013}.

The mean squared displacement $\langle(\Delta f)^2\rangle$ is expected to exhibit a finite size scaling behavior of type
\begin{eqnarray}
\langle(\Delta f)^2\rangle\sim\delta+\alpha L^{2-\eta}\cdot \left(1+\zeta(L)\right)\,.
\label{eqn:bchbqnqjnxqjxnqsnbqdqwinaiv}
\end{eqnarray}
The main obstacle to overcome in an attempt to determine the exponent $\eta$ with high precision is to 
assess the factor $\zeta(L)$ which hosts the subleading corrections to scaling by constructing an appropriate ansatz. 
In principle, these corrections arise from the presence of irrelevant couplings, and thus the leading contributions to
$\zeta(L)$ should correspond to powers of $1/L^{\om},1/L^{\om_2}, \dots$, where $\om_2$ denotes the Wegner exponent of the next-to-leading
irrelevant coupling \cite{Barber_FSS_1983,Privman_FSS_1990} (in Ref.~\cite{Troester_PRB_87_104112_2013}, an additional logarithmic contribution of type
$\zeta(L)=\beta \ln L+\gamma/L^\om+\dots$ has already been ruled out).
Unfortunately, however, we had (and have) been unable to spot any published numerical estimate for the correction to scaling exponent $\omega$ in the literature. 
Interestingly, while such an estimate may have been beyond reach for previous simulation approaches to the flat phase of tethered membranes,
it seems as if $\om$ is equally hard to extract from analytical methods \cite{Wiese_PMAR_2000}. 
Lacking any estimate of the correction to scaling exponent $\omega$, in Ref.~\cite{Troester_PRB_87_104112_2013} 
we had chosen to monitor the error bars for the remaining fit parameters produced by different choices of 
$\om$ in the interval $[0,1]$.  Based on this reasoning, the estimate $\eta = 0.795(10)$ of Ref.~\cite{Troester_PRB_87_104112_2013} quoted above
had finally been derived for the ``naive'' choice  $\zeta(L)=\beta/L+\gamma/L^2$. Unfortunately, however,
the sign of the resulting value $\delta$ produced in this fit is positive, a fact that was not paid much attention to in Ref.~\cite{Troester_PRB_87_104112_2013}.
Recall that according to Eqn.~(5) of Ref.~\cite{Troester_PRB_87_104112_2013}
\begin{eqnarray}
\langle(\Delta f)^2\rangle  \sim \int\frac{d^2q}{(2\pi)^2}\tilde G(\bq)
\label{eqn:cdhvbhbhxbhbqh}
\end{eqnarray}
where asymptotically for $|\bq|\to0$
\begin{eqnarray}
\tilde G(\bq)=\langle|\tilde f(\bq)|^2\rangle\sim\frac{1}{\kappa q^{4-\eta}}  
\end{eqnarray}
Based on this ideal power law, a spherical cutoff geometry $2\pi/L\le |\bq|\le \La$ 
results in 
\begin{eqnarray}
(\Delta f)^2  \sim
\int_{2\pi/L}^{\La} \frac{dk/(2\pi)^2}{\kappa k^{3-\eta}}
=\frac{-\La^{2-\eta}+(L/2\pi)^{2-\eta}}{(2\pi)^2(2-\eta)\kappa}
\end{eqnarray}
which implies a \emph{negative} value $\delta=-\La^{2-\eta}/(2\pi)^2(2-\eta)\kappa$, and
it is reasonable to expect that this heuristic observation also carries over to the case of cubic cutoff geometry with subleading scaling corrections included.

\begin{figure}[tbp]
\centering
\includegraphics[width=\columnwidth]{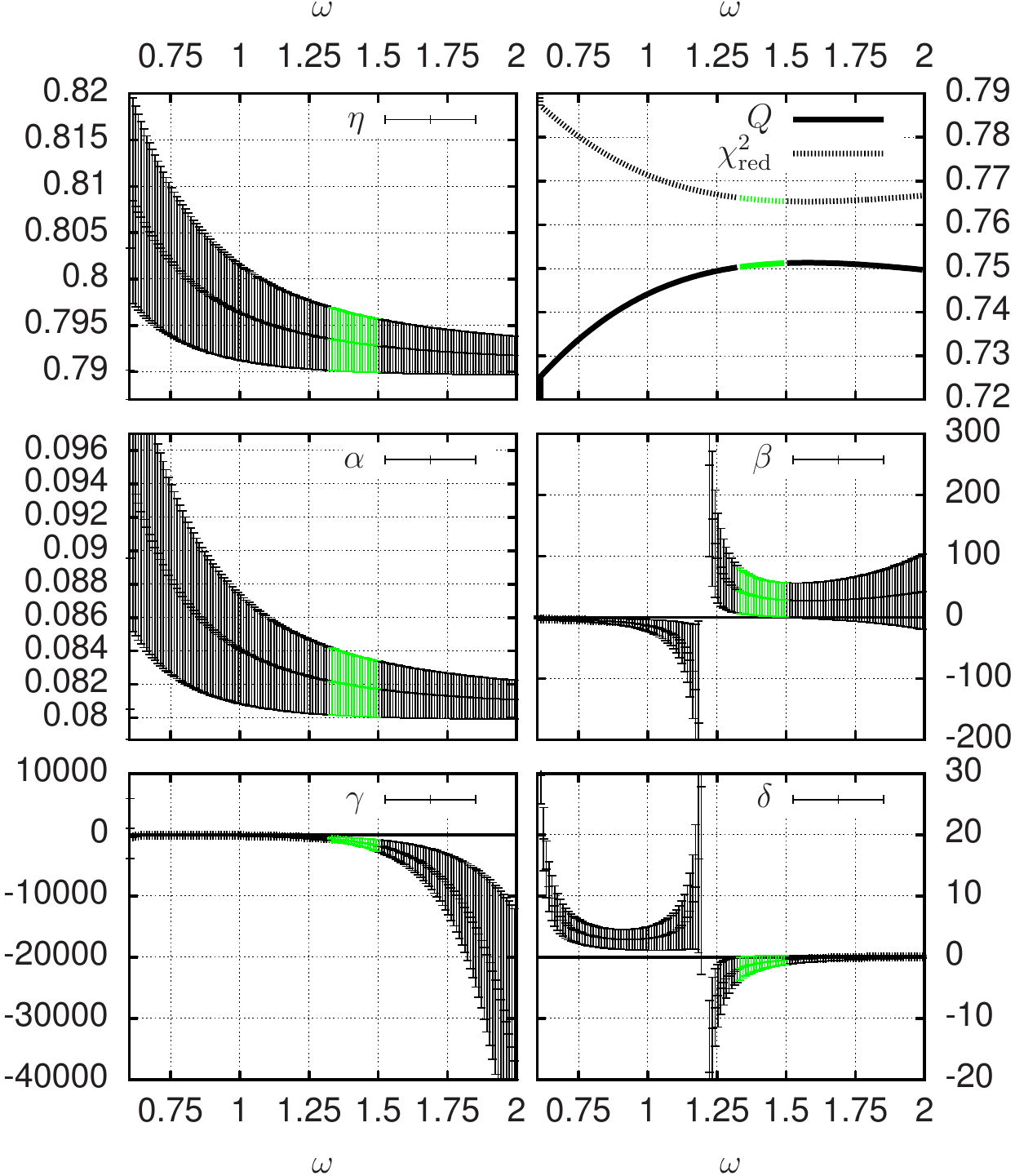}
\caption{Results of least-squares fits of the ansatz (\ref{eqn:bchbqnqjnxqjxnqsnbqdqwinaiv}) to the data for $\langle(\Delta f)^2\rangle$
obtained in Ref.~\cite{Troester_PRB_87_104112_2013} for various choices $0.6\le\om\le2.0$. Data within the range $4/3\le\om\le3/2$ are plotted in green color
as a guide to the eye.
Top left panel: fit results for $\eta$.
Top right panel: $\om$-dependence of goodness-of-fit parameter $Q$ and reduced $\chi^2$-parameter of the fits.
Bottom right panel:  $\om$-dependence of fit parameter $\delta$. Note the discontinuous change of sign around $\om\approx 1.22$.
Remaining panels:  $\om$-dependence of other fit parameters $\al,\beta$ and $\gamma$.
}
\label{fig:OmegaDependenceNew}
\end{figure}

Our present RG approach, in which $\om(b)$ is derived in (\ref{eqn:djdnejdnjendjnejnjednwednjqwnd}) as a by-product of locating the 
FP coupling $K^*(b)$ and the exponent $\eta(b)$ without extra effort, now puts us in a position to shed some new light on these problems, 
even though some residual speculations on the structure of $\zeta(L)$ are still involved. We propose the ansatz 
\cite{Barber_FSS_1983,Privman_FSS_1990,AmitMartinMayor_FRRGCP_2005,PelissettoVicari_PR368_549_2002,Hasenbusch_PRB82_174433_2010}
\begin{eqnarray}
\zeta(L)=\beta/L^\om+\gamma/L^{2\om}+\dots
\label{eqn:bchbqnqjnxqjxnqsnbqdqwi}
\end{eqnarray}
which amounts to assuming that either $\om_2\approx 2\om$ or $\om_2\gg 2\om$, and also to completely discarding additional ``analytic'' corrections
of type $1/L$. While the first assumption is admittedly difficult to justify based on our present knowledge,
the latter appears to be reasonable for periodic boundary conditions where the renormalization of the scaling field $g_L=1/L$ is trivial \cite{SalasSokal_cond-mat/9904038v2}.
To evaluate the impact of these scenarios on the numerical estimate of $\eta$, we have carried out
new fits of the FSS ansatz (\ref{eqn:bchbqnqjnxqjxnqsnbqdqwinaiv}) for various choices $0.6\le\om\le2.0$ 
to the data for $\langle(\Delta f)^2\rangle$ generated in  Ref.~\cite{Troester_PRB_87_104112_2013}. 
The results, which are shown in Fig.~\ref{fig:OmegaDependenceNew}, reveal a number of interesting points.
According to Fig.~\ref{fig:OmegaDependenceNew}, $\delta<0$  appears to hold only for $\om$ somewhat larger than $2-\eta\approx 1.22$, 
a range of values for $\om$ that had unfortunately
not been considered in Ref.~\cite{Troester_PRB_87_104112_2013}. Our fits obviously become singular near this value, but this should not come
as a surprise: quite trivially, for $\om$ equal to $2-\eta$ multiplication of the correction  $\beta/L^\om$ in (\ref{eqn:bchbqnqjnxqjxnqsnbqdqwi}) with $L^{2-\eta}$ produces
yet another constant besides $\delta$, which results in an ill-defined fitting prescription in the close vicinity of this value of $\om$.
As the top left plot of Fig.~\ref{fig:OmegaDependenceNew}) indicates, for both our two possible extrapolations $\om=4/3$ and $\omega=3/2$, the parameter $\delta$ is indeed negative.
We obtain
\begin{eqnarray}
\eta = \entwederoder{0.7935(34)}{\om=4/3}{0.7927(29)}{\om=3/2}
\label{eqn:jnnssddncsdncsdjcndjcn}
\end{eqnarray}
The accompanying goodness-of-fit parameter $Q$ \cite{Press_NR_2007} 
displays a minimum near $\om=3/2$ (cf.~the top right plot of Fig.~\ref{fig:OmegaDependenceNew}), thus slightly 
favoring the second of the two scenarios (\ref{eqn:jnnssddncsdncsdjcndjcn}).   
In retrospective, we note that their common denominator, namely the assertion $\om>1$, 
could have been already anticipated from a FSS analysis similar to the carried out in Ref.~\cite{Troester_PRB_87_104112_2013}
if only we had monitored the sign of the resulting fit parameter $\delta$ in the ansatz (\ref{eqn:bchbqnqjnxqjxnqsnbqdqwinaiv}) for $\langle(\Delta f)^2\rangle$.

In closing this section we note that our new asymptotic fitting procedure, which we have used above to compute coarse-grained parameters
and which is explained in the Appendix, offers yet a complementary way to estimate $\eta$ from the FSS of $\langle(\Delta f)^2\rangle$.
After all, it was designed for the very purpose of extracting leading functional dependencies from data with unknown higher order corrections.
Based on application of the linear ansatz $\log\langle(\Delta f)^2\rangle\sim \log\alpha +(2-\eta)\log L$ to the logarithms of the data points,
the result $\eta=0.7948$ obtained by our simple recipe
is impressively close to that of our above elaborate FSS analysis, with nothing more than the leading scaling behavior as input,
even though it may be difficult to estimate the corresponding error bar.

\section{Discussion and Outlook}

In this paper we have illustrated the practical feasibility and usefulness of our FMC implementation of Wilson's 
MSRG for the flat phase of crystalline membrane, a nontrivial model of continuing physical interest. In particular, we have demonstrated the ability
of our method to derive the - albeit crude - numerical estimate $\om\pm0.3\in[4/3,3/2]$ for the correction to scaling exponent in a situation where all other 
approaches have failed so far. Yet, for a meaningful numerical analysis it is mandatory to  
monitor the dependence of observables on the thickness parameter $b$ of the employed momentum shell.

Our RG result for $\omega$ also allowed to construct an improved FSS procedure for $\langle(\Delta f)^2\rangle$. The resulting  
new estimates (\ref{eqn:jnnssddncsdncsdjcndjcn}) for the exponent $\eta$  deviate even more from the value $\eta=0.85$ derived
both from functional RG \cite{KownackiMouhanna_PRE79_040101_2009,Braghin_PRB82_035407_2010,Hasselmann_PRE83_031137_2011} 
than our previous one $\eta = 0.795(10)$ given in Ref.~\cite{Troester_PRB_87_104112_2013}. In fact, they happen to
be much closer to the result $\eta=0.78\,22(5)$ extracted from a second order self-consistent screening approximation \cite{Gazit_PRE80_041117_2009}.
On the other hand, even our own RG results (\ref{eqn:chbbchcchcbcbbw}) for $\eta$ show a similar tendency to exceed the FSS estimates. 
A heuristic explanation for this common tendency of RG approaches to overestimate $\eta$ is as follows.

For the flat membrane model, Fig.~\ref{fig:ergvonb} indicates that 
the influence of irrelevant couplings is minimal for $b\to\infty$. Still, even in this limit 
our RG estimate (\ref{eqn:chbbchcchcbcbbw}) is noticeably higher than all those obtained from FSS, no matter which 
kind of scaling corrections we employ, even though the underlying data were generated
using the same underlying FMC algorithm. As Fig.~\ref{fig:ergvonb} indicates, our RG analysis seems not to be afflicted with appreciable
finite size effects. Thus, the only explanation for this discrepancy is a residual systematic error in the RG result due to the influence of the 
irrelevant couplings that survives the limit $b\to\infty$. In fact, had we not been carefully monitoring the $b$-dependence of our results but
had simply chosen one particularly convenient shell configuration, our result for $\eta$ might have been still dramatically higher, as 
the middle panel of Fig.~\ref{fig:ergvonb} indicates. In view of the fact that 
generic functional RG calculations also include choosing a projection to a low-dimensional coupling constant space and a cutoff function,
these observations may hint at the source of the persistent discrepancy 
between RG and FSS estimates for $\eta$.

In the near future we plan to investigate the critical behavior of hexatic membranes in the 
so-called crinkled phase \cite{DavidGuitterPeliti_JP48_1987,BowickTravesset2001,GuitterKardar_EPL13_1990,CodelloZanusso_PRE88_022135_2013} using a similar strategy.
Compared to that of crystalline membranes, this problem is closely related but technically much more involved
due to the fact that for hexatic membranes a surface tension contribution of type
\begin{eqnarray}
\calH^{\La,s}=\mu\int\frac{d^2q}{(2\pi)^2}q^2|\tilde f(\bq)|^2
\label{eqn:xjxncnjcjncjcnjcjdcnjdcjdnjdcnc}
\end{eqnarray}
has to be taken into account, its coupling constant $\mu$ being relevant in the RG sense \cite{CodelloZanusso_PRE88_022135_2013}. 
Using conventional FMC we have found it very difficult to determine the critical value $\mu_c^*(\kappa)$ to which $\mu$ must be tuned 
at a given value of $\kappa$ to actually observe the crinkled phase  
\cite{Troester_JPCS510_012008_2014}. Our present approach offers a way to do this, but at the expense of determining a two-dimensional flow pattern in 
the variables $\kappa$ and $\mu$ at fixed parameter $K$. Work in this direction is currently in progress. 

\appendix*
\section{Appendix}
To demonstrate the fitting strategy used in extracting the lowest order dispersion coefficients from the functions $a(k)$ and $b(k)$,
we consider a simple toy model defined by the function
\begin{eqnarray}
f(x)=0.09 x^2-0.1 x^4-0.2 x^6+0.9 x^8
\label{eqn:cbqqhdbqhdbqhdbqhdqhwdqwdbqw}
\end{eqnarray}
in the interval $[0,1]$. We choose the $20$ equidistant points $x_n:=0.05\cdot n$, $n=1,\dots 20$ from this interval,
and generate the telescopic data sets
\begin{eqnarray}
F_k:=\{(x_n,f(x_n)):n=1,\dots,k\},\ k=1,\dots,20 \,.\nonumber
\\&&
\label{eqn:gdgdggdgbddaysxgvscvhgxcvasjcb}
\end{eqnarray}
We will use the largest considered data set $F_{20}$ holding $20$ function values as the input data to our
procedure, whose goal it is to reconstruct the leading coefficient $0.09$ of $f(x)$,
based only on the information that $f(x)$ should start out 
with a term $\propto x^2$ multiplied by a non-negative coefficient. 

We start by choosing the fit function 
\begin{eqnarray}
\ph(x):=a^2x^2 
\label{eqn:hcbwhbchwbbhwebhwebhewbhewdbhewbd}
\end{eqnarray}
which deliberately ignores all higher order corrections to this leading $x$-dependence that should gradually
kick in for growing values of $x$. Of course, the quality of a series of least-squares fits of this function applied to the
sets $F_k$ will successively degrade with growing $k$. Quantitatively, we will observe a crossover from a slow to a steep rise of the accompanying $\chi^2$-parameters 
\begin{eqnarray}
\chi_k^2:=\sum_{l=1}^{k}[f(x_l)-\ph(x_l)]^2,\ k=1,\dots 20
\end{eqnarray}
with growing $k$. The idea is to use the inverse values 
\begin{eqnarray}
w_k:=\entwederoder{1/\chi_2^2}{k\le2}{1/\chi_k^2}{k>2}
\label{eqn:wewebwhbhwbhwbwhbhwbhwbhwbwhb}
\end{eqnarray}
(since trivially $\chi_1^2$ should vanish, we have put $w_1\equiv w_2$ in (\ref{eqn:wewebwhbhwbhwbwhbhwbhwbhwbwhb}))
as statistical weights for the data points $(x_n,f(x_n)$ in a final least-squares fit 
of (\ref{eqn:hcbwhbchwbbhwebhwebhewbhewdbhewbd}) to the full data set $F_{20}$,
which produces our final estimate for $a^2$.

\begin{figure}[tbp]
\centering
\includegraphics[scale=0.6]{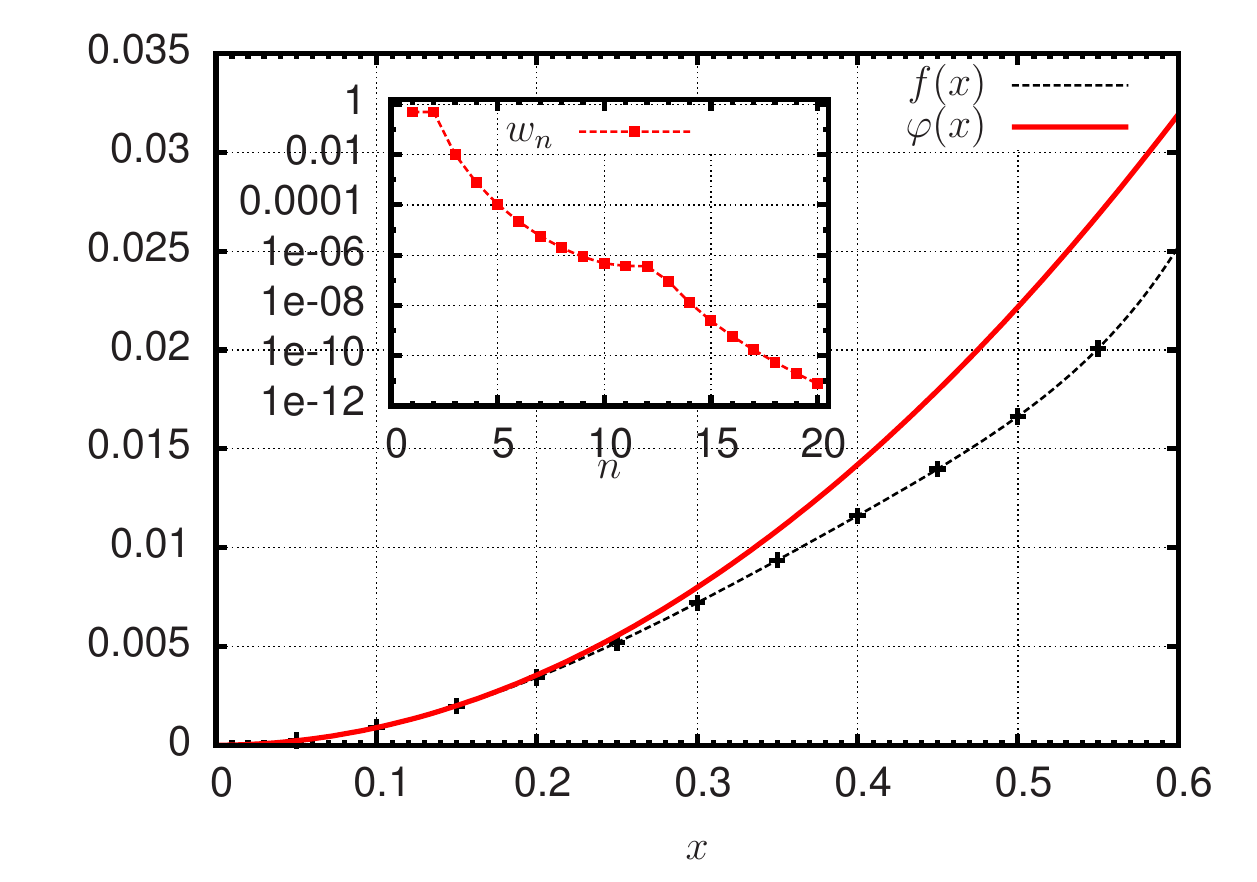}
\caption{Illustration of the fit procedure outlined in Eqs.~(\ref{eqn:cbqqhdbqhdbqhdbqhdqhwdqwdbqw})-(\ref{eqn:wewebwhbhwbhwbwhbhwbhwbhwbwhb}).
Main plot: original data set $F_{20}$ produced from the function (\ref{eqn:cbqqhdbqhdbqhdbqhdqhwdqwdbqw}) and final fit with Eqn.~(\ref{eqn:hcbwhbchwbbhwebhwebhewbhewdbhewbd})
(but only shown in the interval $(0,0.6)$). 
Inset: normalized weights obtained for all $20$ data points of the data set $F_{20}$ (note the logarithmic scale).   
}
\label{fig:fitstudy}
\end{figure}

The described procedure is simple to implement and robust, but, of course, far from perfect.
Obviously, its success relies on a high quality of the underlying data. In particular, it is vulnerable
to statistical outliers located inside the asymptotic region $x\ll 1$. Nevertheless, at least in the context of the present paper,
where the unknown higher order correction terms 
can be expected to be small compared to the leading term, it seems to produce sufficiently accurate results.
For example, in the case of the toy function (\ref{eqn:cbqqhdbqhdbqhdbqhdqhwdqwdbqw}), we obtain $a=0.29781$, which is
within $0.73\%$ of the exact value $0.3=\sqrt{0.09}$. 
In view of the fact that we did not have to make any assumptions on the structure of the fitting function
beyond its lowest order term, the achieved precision is quite satisfactory. 
Fig.~\ref{fig:fitstudy} illustrates the resulting final fit of $\ph(x)$ to the full data set $F_{20}$ obtained by reweighting its data using the weights 
$w_k$ successively generated by the previous fits to the data sets $F_k$.

\begin{acknowledgments}
We appreciate valuable advice from M. Hasenbusch concerning our finite size scaling approach.
We acknowledge support by the Austrian Science Fund (FWF) Project P22087-N16.
Major parts of our computations were performed on the Vienna Scientific Cluster (VSC1 and VSC2). 
\end{acknowledgments}



\begin{thebibliography}{10}

\bibitem{Wilson_RMP55_1983}
K.~G. Wilson, Rev. Mod. Phys. {\bf 55},  583  (1983).

\bibitem{Zinn-Justin_QFTCP_2002}
J. Zinn-Justin, {\em Quantum Field Theory and Critical Phenomena} (Oxford Univ.
  Press, Oxford, 2002).

\bibitem{KleinertSchulteFrolinde_CP_2001}
H. Kleinert and V. Schulte-Frohlinde, {\em Critical Properties of
  $\phi^4$-Theories} (World Scientific, Singapore, 2001).

\bibitem{Kopietz2010}
P. Kopietz, L. Bartosch, and F. Sch{\"u}tz, {\em Introduction to the Functional
  Renormalization Group} (Springer, Berlin Heidelberg, 2010).

\bibitem{AmitMartinMayor_FRRGCP_2005}
D. Amit and V. Martin-Mayor, {\em Field Theory, the Renormalization Group, and
  Critical Phenomena: Graphs to Computers (3rd Edition)} (World Scientific,
  Singapore, 2005).

\bibitem{LandauBinder_MC_2009}
D. Landau and K. Binder, {\em A Guide to Monte Carlo Simulations in Statistical
  Physics}, 3rd  ed. (Cambridge University Press, Cambridge, 2009).

\bibitem{Troester_PRB76_2007}
A. Tr{\"o}ster, Phys. Rev. B {\bf 76},  012402  (2007).

\bibitem{Troester_CSSCMP_2008}
A. Tr{\"o}ster and C. Dellago, Physics Procedia {\bf 6},  106   (2010).

\bibitem{Troester_PRL100_2008}
A. Tr{\"o}ster, Phys. Rev. Lett. {\bf 100},  140602  (2008).

\bibitem{TroesterCPC179_2008}
A. Tr{\"o}ster and Computer Physics Communications {\bf 179},  30 (2008).

\bibitem{Troester_PRB_81_012406_2010}
A. Tr\"oster, Phys. Rev. B {\bf 81},  012406  (2010).

\bibitem{Troester_PRE_79_036707_2009}
A. Tr\"oster, Phys. Rev. E {\bf 79},  036707  (2009).

\bibitem{Troester_CPC182_2011}
A. Tr{\"o}ster, Comput. Phys. Commun. {\bf 182},  1837   (2011).

\bibitem{BruceDrozAharony_JPCSSP7_1974}
A. Bruce, M. Droz, and A. Aharony, J. Phys. C: Solid State Physics {\bf 7},
  3673  (1974).

\bibitem{Troester_PRB81_125135_2010}
A. Tr\"oster, Phys. Rev. B {\bf 81},  125135  (2010).

\bibitem{FisherMaNickel_PRL_1972}
M. Fisher, S. Ma, and B. Nickel, Phys. Rev. Lett. {\bf 29},  917  (1972).

\bibitem{NelsonPiranWeinberg_Membranes_1988}
{\em Statistical Mechanics of Membranes and Surfaces}, edited by D. Nelson, T.
  Piran, and S. Weinberg (World Scientific, Singapore, 1988).

\bibitem{Troester_PRB_87_104112_2013}
A. Tr\"oster, Phys. Rev. B {\bf 87},  104112  (2013).

\bibitem{LeDoussalRadzihovsky_PRL69_1992}
P. {Le Doussal} and L. Radzihovsky, Phys. Rev. Lett. {\bf 69},  1209  (1992).

\bibitem{Gazit_PRE80_041117_2009}
D. Gazit, Phys. Rev. E {\bf 80}, 041117 (2009).

\bibitem{AronovitzLubensky_PRL60_1988}
J.~A. Aronovitz and T.C. Lubensky, Phys. Rev. Lett. {\bf 60},  2634  (1988).

\bibitem{DavidGuitter_EPL5_1988}
F. David and E. Guitter, Europhys. Lett. {\bf 5},  709  (1988).

\bibitem{KownackiMouhanna_PRE79_040101_2009}
J.-P. Kownacki and D. Mouhanna, Phys. Rev. E {\bf 79},  040101  (2009).

\bibitem{Braghin_PRB82_035407_2010}
F.~L. Braghin and N. Hasselmann, Phys. Rev. B {\bf 82},  035407  (2010).

\bibitem{Hasselmann_PRE83_031137_2011}
N. Hasselmann and F.~L. Braghin, Phys. Rev. E {\bf 83},  031137  (2011).

\bibitem{Bowick_JPF6_1321_1996}
M.~J. Bowick, S.~M. Catterall, M. Falcioni, G. Thorleifsson, and K.~N.
  Anagnostopoulos, J. Phys. I (France) {\bf 6},  1321  (1996).

\bibitem{LosFasolino_PRB80_121405_2009}
J.~H. Los, M.~I. Katsnelson, O.~V. Yazyev, K.~V. Zakharchenko, and A. Fasolino,
  Phys. Rev. B {\bf 80},  121405(R)  (2009).

\bibitem{Wiese_PMAR_2000}
J.~K. Wiese,  in {\em Phase Transitions and Critical Phenomena}, edited by
  C.Domb and J.Lebowitz (Academic Press, London, 2000), Chap.~Polymerized
  Membranes, a Review.

\bibitem{Kleinert_PathIntegrals_2009}
H. Kleinert, {\em Path Integrals in Quantum Mechanics, Statistics, Polymer
  Physics, and Financial Markets} (World Scientific, Singapore, 2009).

\bibitem{Luijten_PhD_1997}
E. Luijten, Ph.D. thesis, Delft University of Technology, The Netherlands,
  1997.

\bibitem{Safran_STSIM_2003}
S.~A. Safran, {\em Statistical Thermodynamics of Surfaces, Interfaces, and
  Membranes} (Perseus Books, Cambridge MA, USA, 2003).

\bibitem{Katsnelson_Graphene_2012}
M.~I. Katsnelson, {\em Graphene: Carbon in Two Dimensions} (Cambridge
  University Press, Cambridge, UK, 2012).

\bibitem{TroesterDellago_F354_2007}
A. Tr{\"o}ster and C. Dellago, Ferroelectrics {\bf 354},  225  (2007).

\bibitem{Troester_PP53_496_2014}
A. Tr{\"o}ster, Physics Procedia {\bf 53},  96  (2014).

\bibitem{Wegner_PRB5_4529_1972}
F.~J. Wegner, Phys. Rev. B {\bf 5},  4529  (1972).

\bibitem{Belardinelli_JCP127_2007}
R.~E. Belardinelli and V.~D. Pereyra, J. Chem. Phys. {\bf 127},  184105
  (2007).

\bibitem{Belardinelli_PRE_2008}
R.~E. Belardinelli, S. Manzi, and V.~D. Pereyra, Phys. Rev. E {\bf 78},  067701
   (2008).

\bibitem{WangLandau_PRL_2001}
F. Wang and D.~P. Landau, Phys. Rev. Lett. {\bf 86},  2050  (2001).

\bibitem{Landau_BrazJPhys34_354_2004}
D.~P. Landau and F. Wang, Braz. J. Phys. {\bf 34},  354   (2004).

\bibitem{EfronTibshirani1994}
B. Efron and R. Tibshirani, {\em An Introduction to the Bootstrap} (Chapman and
  Hall, Boca Raton, 1994).

\bibitem{Barber_FSS_1983}
M. Barber,  in {\em Phase Transitions and Critical Phenomena}, edited by C.
  Domb and J. Lebowitz (Academic Press, New York, 1983), Chap.~2. Finite Size
  Scaling, p.\ 145.

\bibitem{Privman_FSS_1990}
V. Privman, {\em Finite Size Scaling and Numerical Simulation of Statistical
  Systems} (World Scientific, Singapore, 1990).


\bibitem{PelissettoVicari_PR368_549_2002}
A. Pelissetto and E. Vicari, Phys. Rep. {\bf 368}, 549 (2002).

\bibitem{Hasenbusch_PRB82_174433_2010}
M. Hasenbusch, Phys. Rev. B {\bf 82}, 174433 (2010).

\bibitem{SalasSokal_cond-mat/9904038v2}
J. Salas and A.D. Sokal, cond-mat/9904038v2 (1999). 
 
\bibitem{Press_NR_2007} W.H. Press, S.A. Teukolsky, W.T. Vetterling, and B.P. Flannery,
{\em Numerical Recipes: The Art of Scientific Computing}, 3rd Ed.
(Cambridge Univ. Press, Cambridge, UK, 2007).

\bibitem{DavidGuitterPeliti_JP48_1987}
F. David, E. Guitter, and L. Peliti, J. Physique {\bf 48},  2059  (1987).

\bibitem{BowickTravesset2001}
M. Bowick and A. Travesset, Phys. Rep. {\bf 344},  255  (2001).

\bibitem{GuitterKardar_EPL13_1990}
E. Guitter and M. Kardar, Europhys. Lett. {\bf 13},  441  (1990).

\bibitem{CodelloZanusso_PRE88_022135_2013}
A. Codello and O. Zanusso, Phys. Rev. E {\bf 88},  022135  (2013).

\bibitem{Troester_JPCS510_012008_2014}
A. Tr{\"o}ster, Journal of Physics: Conference Series {\bf 510},  012008
  (2014).

\end{thebibliography}


\end{document}